\documentclass[fleqn,usenatbib]{mnras}
\usepackage{mathptmx}
\usepackage{newtxtext,newtxmath}
\usepackage{longtable}
\usepackage[T1]{fontenc}
\usepackage{ae,aecompl}

\usepackage{graphicx}	
\usepackage{booktabs, dcolumn}
\usepackage{pdflscape}
\usepackage{afterpage}
\usepackage[usenames,dvipsnames]{xcolor}

\newcommand{\angstrom}{\textup{\AA}}
\newcommand{\Teff}{\mbox{$T_{\mathrm{eff}}$}}
\newcommand{\Msun}{\mbox{$\mathrm{M_{\odot}}$}}
\newcommand{\Rsun}{\mbox{$\mathrm{R_{\odot}}$}}

\title[White dwarfs with planetary remnants in Gaia I]{White dwarfs with planetary remnants in the era of \textit{Gaia} I:\\ six emission line systems}

\author[Gentile Fusillo et al.]{N.~P.~Gentile Fusillo$^{1}$\thanks{E-mail: ngentile@eso.org}, C. J. Manser$^{2}$, Boris~T.~G\"ansicke$^{2}$, O. Toloza $^{2}$, D.~ Koester$^3$, \newauthor E.~Dennihy$^4$,
W.~R.~Brown$^{5}$, J. Farihi$^{6}$,  M.~A.~Hollands$^{2}$, M.~J.~Hoskin$^2$, \newauthor P.~Izquierdo$^{7,8}$, T.~Kinnear$^{9}$, 
 T.~R.~Marsh$^{2}$, A.~Santamar\'ia-Miranda$^{10}$, A.~F.~Pala$^{1}$,  \newauthor S. Redfield$^{11}$, P.~Rodríguez-Gil$^{7,8}$,  M.~R.~Schreiber$^{12,13}$, Dimitri~Veras$^{2}$\thanks{STFC Ernest Rutherford Fellow}, D.~J.~Wilson$^{14}$
 \newauthor
\\
$^{1}$European Southern Observatory, Karl Schwarzschild Stra{\ss}e 2, Garching, 85748, Germany
\\
$^{2}$Department of Physics, University of Warwick, Coventry, CV4 7AL, UK\\
$^{3}$ Institut f\"ur Theoretische Physik und Astrophysik, University of Kiel, 24098 Kiel, Germany\\
$^4$ Gemini Observatory/NSF's NOIRLab, Casilla 603, La Serena Chile\\
$^{5}$Harvard \& Smithsonian $|$ Center for Astrophysics, 60 Garden Street,Cambridge, MA 02138 USA\\
$^{6}$Physics and Astronomy, University College London, London, WC1E 6BT, UK\\
$^{7}$Instituto de Astrof\'isica de Canarias, Calle V\'ia L\'actea, 38205 San Crist\'obal de La Laguna,
Spain\\
$^{8}$Departamento de Astrof\'isica, Universidad de La Laguna, 38206 La Laguna, Tenerife, Spain\\
$^{9}$School of Physical Sciences, Ingram Building University of Kent, Canterbury, CT2 7NH, UK\\
$^{10}$European Southern Observatory, Av. Alonso de C\'ordova 3107, Casilla 19001, Santiago, Chile\\
$^{11}$Department of Astronomy and Van Vleck Observatory, Wesleyan University, Middletown, CT 06459, USA\\
$^{12}$Departamento de F\'isica, Universidad T\'ecnica Federico Santa Mar\'ia, Av. España 1680, Valpara\'iso, Chile\\
$^{13}$N\'ucleo Milenio Formaci\'on Planetaria - NPF, Av. España 1680, Valpara\'iso, Chile \\
$^{14}$McDonald Observatory, University of Texas at Austin, Austin, TX 78712, USA}

\date{Accepted XXX. Received YYY; in original form ZZZ}

\pubyear{2020}
\begin{document}
\label{firstpage}
\pagerange{\pageref{firstpage}--\pageref{lastpage}}
\maketitle

\begin{abstract}
White dwarfs with emission lines from gaseous debris discs are among the rarest examples of planetary remnant hosts, but at the same time they are key objects for studying the final evolutionary stage of planetary systems. Making use of the large number of white dwarfs identified in \textit{Gaia} DR2, we are conducting a  survey of planetary remnants and here we present the first results of our search: six  white dwarfs with gaseous debris discs. This first publication focuses on the main observational properties of these objects and highlights their most unique features. Three systems in particular stand out: WD\,J084602.47+570328.64 displays an exceptionally strong infrared excess which defies the standard model of a geometrically-thin, optically-thick dusty debris disc; WD\,J213350.72+242805.93 is the hottest gaseous debris disc host known with $\Teff=29\,282$\,K; and WD\,J052914.32--340108.11, in which we identify a record number of 51 emission lines from five elements. These discoveries shed light on the underlying diversity in gaseous debris disc systems and bring the total number of these objects to 21. With these numbers we can now start looking at the properties of these systems as a class of objects rather than on a case-by-case basis.

\end{abstract}

\begin{keywords}
white dwarfs -- Circumstellar matter -- planetary systems -- line: profiles --  stars: individual:WD\,J023415.51--040609.28 -- stars: individual: WD\,J052914.32--340108.11 -- stars: individual:WD\,J084602.47+570328.64 -- stars: individual: WD\,J193037.65--502816.91 -- stars: individual:WD\,J213350.72+242805.93 -- stars: individual: WD\,J221202.88--135240.13
\end{keywords}

\section{Introduction}
Over 95\,per\,cent of all stars in the Galaxy, including our Sun and virtually all known planet hosts, will end their lives as white dwarfs. It is now well established that planets can survive the late evolution of their host star \citep{villaver+livio07-1, mustill+villaver12-1, raoetal18-1, martinetal20-1, vanderburgetal20-1}, and evidence of such planetary remnants is seen in the photospheric metal contamination of a large number white dwarfs \citep{zuckermanetal10-1, koesteretal14-1, hollandsetal17-1, barstowetal14-1, schreiberetal19-1}. Heavy element pollution in the atmosphere of white dwarfs is largely the result of the accretion of rocky planetesimals which have been scattered into the tidal disruption radius of the white dwarf, where they are torn apart into dusty debris that subsequently spreads out to form a circumstellar disc \citep{verasetal14-1, verasetal15-1, redfieldetal17-1,  malamud+perets20-1, malamud+perets20-2}. These debris discs can be detected via reprocessed infrared (IR) flux in excess to what is expected from a single isolated white dwarf. About 1--3\,per\,cent of all white dwarfs display this IR signature  \citep{barberetal12-1, rocchettoetal15-1, wilsonetal19-1, rebassa-mansergasetal19-1}. In addition to the thermal IR emission from the dust, a small subset of these already rare systems \citep{manseretal20-1} also display line emission typically strongest at the Ca\,{\textsc{ii}} 8600\,\AA\ triplet due to the presence of a gaseous component of the disc \citep{gaensickeetal06-3} . These Doppler-broadened, emission features are largely consistent with models for gas in Keplerian orbits in a flat disc \citep{horne+marsh86-1}, though evidence indicates that these orbits can reach high eccentricities. \citep{gaensickeetal08-1,cauleyetal18-1, manseretal19-1}. In a large fraction of these systems the Ca emission features show some level of asymmetry, suggesting a non-axisymmetric intensity distribution in the disc. Long term monitoring of the prototype SDSS\,J1228+1040 has revealed a smooth evolution of the line profile morphology interpreted as the precession of a persisting asymmetric intensity pattern with a period of $\simeq27$\,yr \citep{manseretal16-1}. A similar morphological evolution of emission lines has also been observed in the gaseous debris disc around HE\,1349--2305 \citep{dennihyetal18-1}, albeit on a much shorter period of $\simeq$\,1.4\,yr. Where information on the spatial distribution of the gas is available, it appears to be co-located with the dust and thus not confined within the sublimation radius of the dust \citep{melisetal10-1, manseretal16-1}. These findings imply that some mechanism keeps generating the gas which would otherwise re-condense on time scales of months \citep{metzgeretal12-1}. 
The exact process underlying the generation of the circumstellar gas remains intensely debated and different scenarios  have been proposed, including dust sublimation at the inner edge of the debris disc followed by subsequent radial spreading \citep{rafikov11-2, metzgeretal12-1}, and collisional cascades grinding the debris into gas \citep{kenyon+bromley17-1, kenyon+bromley17-2}.

Recently, a follow-up study of SDSS\,J1228+1040 revealed additional variability in the strength and shape of the gas emission features with a period of 123.4\,min. The authors interpreted this periodic signal as the orbital signature of a dense planetesimal, with sufficient internal strength to survive tidal disruption, embedded in the dust disc \citep{manseretal19-1}. \citet{manseretal19-1, manseretal20-1} also speculated that the presence of such solid planetary bodies may be at the origin of the gaseous discs in all such systems, particularly since dynamical mechanisms exist to embed and circularize planetesimals within the discs \citep{grishinetal19-1,oconnoretal20-1}.

Recently, \citet{gaensickeetal19-1} announced the discovery of the white dwarf WD\,J091405.30+191412.25 (hereafter WD\,J0914+1914),
which hosts a different kind of gaseous disc of planetary origin. This disc presents no observable dusty component, and both the material in the disc and that accreted onto the white dwarf have compositions rich in volatile elements (H, O, and S), consistent with a giant planet atmosphere. We consider this star a member of a separate class, and for the rest of this  paper "gaseous debris disc" and similar expressions will only refer to systems with debris originating from rocky bodies and which display Ca\,{\textsc{ii}} emission lines.

Though our understanding of these gaseous discs is still incomplete, all evidence indicates that they are the signposts of dynamical instabilities and recent disruption events \citep{xuetal14-1,xuetal18-1,swanetal20-1}, making these systems ideal laboratories to study the formation and evolution of planetary debris discs \citep{rafikov11-2, metzgeretal12-1}, and possibly even the incidence of closely orbiting planetesimals at white dwarfs \citep{manseretal19-1}.
According to recent estimates only 4\,$\pm$\,$_{2}^{4}$\,per\,cent of white dwarfs with dusty debris discs, or 0.067\,$\pm$\,$_{0.025}^{0.042}$\,per\,cent of all white dwarfs display emission features from a gaseous component \citep{manseretal20-1}. The sheer rarity of these objects represents the first and largest obstacle to overcome, but by combining the $\simeq260\,000$ white dwarfs found by \textit{Gaia} \citep{gentilefusilloetal19-1}, with additional infrared data from large area surveys, we now have the opportunity  to find virtually all visible debris disc hosts \citep{rebassa-mansergasetal19-1,dennihyetal20-1,dennihyetal20-2,xuetal20-1}. 

Here, we present the first results from our ongoing search for new white dwarfs with remnants of planetary systems: the identification of six emission line gas disc systems. The work presented here focuses on the discovery of these systems and reports their unique observational properties.
The detailed abundance analysis and modelling of the gaseous discs goes beyond the scope of this paper, and will be presented elsewhere.

\section{A search for IR excess to new white dwarfs}
In our search for remnants of planetary systems we cross-matched all high-confidence white dwarf candidates ($P_{\mathrm{WD}} \geq 0.75$)  brighter than \textit{Gaia} $G=18.5$ from \citet{gentilefusilloetal19-1} with the large area IR surveys: the two micron All Sky Survey (2MASS, \citealt{2mass}), UKIRT Infrared Deep Sky Survey (UKIDSS, \citealt{ukidss}), UKIRT Hemisphere Survey (UHS, \citealt{UHS}, Vista Hemisphere Survey (VHS, \citealt{VHS13-1}), and the Wide-field Infrared Survey Explorer (\textit{WISE}, \citealt{wrightetal10-1}). 

\begin{figure*}
	\includegraphics[width=2\columnwidth]{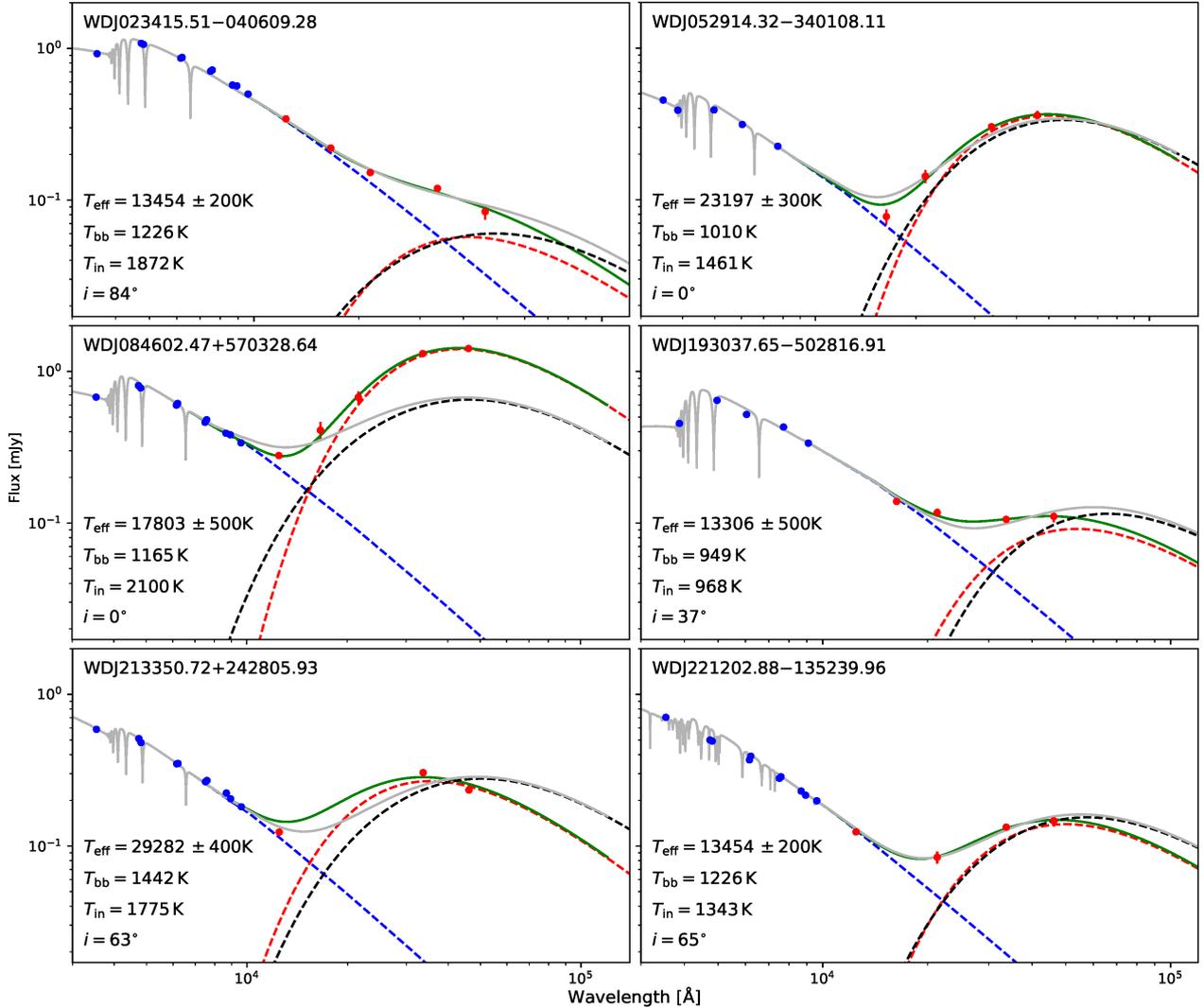}
    \caption{SEDs of the six newly identified gaseous debris disc systems. Optical and IR photometry used for the fits are represented by blue and red dots, respectively (see Table\,\ref{mega_table}).   
    We fitted these SEDs with simple two-component models consisting of a white dwarf spectrum (blue-dashed line) and either a single-temperature blackbody (red dashed line) or the optically-thick, geometrically-thin  debris disc model from \citet[black dashed line]{jura03-1}.  
   The sums of the white dwarf models and blackbodies are displayed as green solid lines; and the sums of white dwarf models and disc models are displayed as grey solid lines. 
   In each panel, we report the disc inner edge temperature,$T_{\mathrm{in}}$, and inclination, $i$, for the statistically best fitting optically-thick, geometrically-thin disc model, but we emphasize that these values should not be considered as a reliable physical description of the geometry of the dusty debris.  
   Parameters of the best-fit white dwarf and blackbody model are also reported in Table\,\ref{mega_table}.}
    \label{SEDs}
\end{figure*}

In order to account for the different epochs of observations across the different surveys, we carried out the cross-match with each survey in separate steps. Firstly, for all \textit{Gaia} white dwarf candidates we retrieved every matching source within a radius of 30\,arcsec in each IR survey. We then computed the difference in epoch of observations for all the matching pairs and ``forward projected'' the  coordinates of all matching sources using \textit{Gaia} proper motions. Finally we repeated the cross-match using a 2\,arcsec matching radius and considered the closest pairs as true matches (see \citealt{gentilefusilloetal17-1} for details). 
We found that $\simeq 62$\,per\,cent of the white dwarfs have reliable \textit{WISE} detections and $\simeq 32$\,per\,cent of them $K_S$  band observation from a near-IR survey.
We then constructed spectral energy distributions (SEDs) for all white dwarf candidates with matching IR photometry. 
\textit{Gaia} optical photometry was fitted with H-atmosphere white dwarf models and the resulting IR flux of the  white dwarf was compared with the IR survey photometry. Targets with at least a 3\,$\sigma$ excess in the \textit{WISE} bands were selected as initial debris disc host candidates (Fig.\,\ref{SEDs}).

The IR coverage provided by \textit{WISE} ($\simeq 33\,500$ \AA\ and $\simeq46\,000$\,\AA) is essential to detect debris disc candidates, but the large beam of the telescope can easily lead to source confusion and thus to false positives (e.g \citealt{dennihyetal20-1}). 
In order to reduce contamination from spurious sources and optimize our spectroscopic follow-up, we prioritize targets which display at least a 2\,$\sigma$ IR-excess already at the near-IR $K_s$ band. The higher spatial resolution of $K_s$ band images facilitates the vetting of potential contamination from background sources, though it cannot fully eliminate this problem \citep{dennihyetal20-1}.
Systems for which the IR flux is in excess only in the \textit{WISE} $W1$ and $W2$ bands were considered lower priority and were observed mostly as backup targets. 
We identified $\simeq 40$ high-confidence debris disc hosts and over 70 additional lower-priority ones. We began an extensive spectroscopic follow-up campaign to confirm the nature of these systems. 
To date we have obtained spectra for  27  systems and, among them, have identified four gaseous debris disc hosts: WD\,J023415.51--040609.28 (WD\,J0234--0406), WD\,J052914.32--340108.11 (WD\,J0529--3401), WD\,J193037.65--502816.91 (WD\,J1930--5028), WD\,J213350.72+242805.93 (WD\,J2133+2428; Fig.\,\ref{SEDs}). WD\,J2133+2428 was observed as a low-priority target (no $K_S$ band observations available) while the other three stars were considered high-confidence targets. All four systems were not known as white dwarfs prior to \textit{Gaia} DR2, highlighting the major impact that of the significantly increased sample of white dwarfs that is available now \citep{gentilefusilloetal19-1}. 

Additionally, we report here on two further systems which were already known before \textit{Gaia}~DR2 and which we identified as debris disc hosts during an earlier spectroscopic follow-up pilot campaign. WD\,J084602.47+570328.64 (WD\,J0846+5703) was first recognised as a H-atmosphere white dwarf (SBSS\,0842+572) by \citet{stepanianetal99-1}. It was recently confirmed to host an IR excess via \textit{Spitzer} observations \citep{swanetal20-1} and it was independently recognised as a gaseous debris disc host by \citet{melisetal20-1}. WD\,J221202.88--135239.96 (WD\,J2212--1352) was first identified as a candidate white dwarf in the ATLAS survey as ATLAS\,J221202.83--135240.13 \citep{gentilefusilloetal17-1}. The IR excess about this star was also confirmed by \textit{Spitzer} observations \citep{dennihyetal20-1}. 

For consistency we refer to all six systems using the \textit{Gaia}-based naming convention WD\,JHHMMSS.SS$\pm$DDMMSS.SS (J2000 coordinates) established in \citet{gentilefusilloetal19-1}, or abbreviation of these names.

\section{Spectroscopic observations of the six gaseous debris hosts}

Spectroscopic observations of the six targets were obtained between 2016 and 2020 using a variety of instruments: X-shooter on the Very Large Telescope (VLT, \citealt{vernetetal11-1}) observatory at Cerro Paranal, Chile;  the Intermediate Resolution Spectrograph \footnote{http://www.ing.iac.es/Astronomy/instruments/isis/} (ISIS) on the William Herschel Telescope (WHT) at Roque de los Muchachos Observatory (ORM) in La Palma, Spain; the Optical System for Imaging and low-Intermediate-Resolution Integrated Spectroscopy (OSIRIS, \citealt{sanchezetal12-1}) on the Gran Telescopio Canarias (GTC) at the ORM; the High Resolution Echelle Spectrometer \cite[HIRES,][]{vogtetal94-1} on the 10-m Keck\,I telescope at Mauna Kea Observatory, Hawaii; the Magellan Echellette Spectrograph (MagE) on the 6.5\,m Baade Magellan telescope at Las Campanas Observatory, Chile; and the FAST spectrograph on the 1.5\,m Fred Lawrence Whipple Observatory telescope on mount Hopkins, Arizona \citep{fabricantetal98-1}.
The observation log is reported in Table\,\ref{OB_log}, and sample spectra are displayed in Fig\,\ref{all_spec} and \ref{Fast_MagE}.

X-shooter observations were carried out using a 1\,arcsec slit aperture for the UVB arm and 0.9\,arcsec for the VIS arm. In all cases flux in the NIR arm was insufficient for any meaningful analysis.
On average the resolving power of our X-shooter spectra is $R\simeq 4000-5500$ in the UVB arm and $R\simeq 6500-8000$ in the VIS arm. All data were reduced using the standard procedures within the {\sc reflex}\footnote{http://www.eso.org/sci/software/reflex/} reduction tool developed by ESO. Telluric lines removal was performed on the reduced spectra using {\sc molecfit} \citep{smetteetal15-1, kauschetal15-1}.

ISIS observations were obtained using the R600B and R600R gratings, in the ISIS blue and red arms respectively; two  central wavelength settings were used for each arm, nominally 3930\,\AA\ and 4540\,\AA\ in the blue and 6562\,\AA\ and 8200\,\AA\ in the red, which were chosen to cover important atmospheric lines, such as the Balmer series and Ca~H and K, as well as the \ion{Ca}{II} emission triplet in addition to other potential emission features (e.g. \ion{Fe}{II}/\ion{Mg}{I} 5175\,\AA).
The slit width varied between 1\,arcsec and 1.5\,arcsec depending on the observing conditions and we employed a binning of $2\times2$, resulting in an average resolution of $\approx2$\,\AA. 
ISIS spectra were de-biased and flat-fielded using the standard {\sc starlink}\footnote{The {\sc starlink} Software Group homepage website is http://starlink.jach.hawaii.edu/starlink.} packages {\sc kappa, figaro} and {\sc convert}. We carried out optimal spectral extraction using the package {\sc pamela}. Wavelength and flux calibration were performed using the routines within the software {\sc molly} \footnote{\label{pamela}{\sc pamela} and {\sc molly} were written by T. R. Marsh and can be found in the {\sc starlink} distribution Hawaiki and later releases.} \citep{marsh89-1}.

The OSIRIS spectroscopy of WD\,J0846+5703 was obtained using the R2500I grism with a 0.6\,arcsec slit, and binning the detector $2\times2$. This setup provides wavelength coverage over $\simeq7335-10\,150$\,\AA\ at a resolution of $R\simeq2500$.  The data were bias and flat-field corrected using the standard tools within {\sc iraf}\footnote{{\sc iraf} is distributed by the National Optical Astronomy Observatories.}. We performed  sky background subtraction and  optimal spectral extraction \citep{horne86-1} using the {\sc pamela} data reduction software \citep{marsh89-1}. Finally, we used {\sc molly}$^{4}$ to wavelength-calibrate the spectra by fitting the arc lamp spectra with low-order polynomials to obtain the pixel-to-wavelength solution.

Additional high-resolution spectroscopy of WD\,J0846+5703 was obtained with HIRES on 2017 May 22. We used the HIRESb configuration with the C5 decker ($1.148 \times 7$\,arcsec slit), covering the spectral range $3100-5950$\,\AA\ and providing a nominal spectral resolving power of $R\approx37\,000$. We retrieved the pipeline-reduced spectra for each individual echelle order from the Keck Observatory Archive, and subsequently combined the data of all three exposures for a few selected emission lines.

FAST and MagE observations were originally conducted for identification of white dwarf candidates, and therefore did not cover the \ion{Ca}{II} triplet. MagE spectra were acquired using a slit width of 1\,arcsec and achieved a resolving power $R \simeq 4100$. The data were reduced using the {\sc python} pipeline based on {\sc CarPy} \citep{kelsonetal00-1, kelson03-1} and flux calibration was then performed using  standard IRAF routines. FAST spectroscopy was obtained using the 600~l/mm grating and the 2\,arcsec slit, which provides $3550-5500$\,\AA\ wavelength coverage at 2.2\,\AA\ spectral resolution. The spectrum was then processed with standard IRAF routines (Mink et al. 2020 AJ submitted).

\begin{center}
\begin{table*}
\caption{\label{OB_log} Log of spectroscopic observations.}
\begin{tabular}{llllll}
\hline
Name& Instrument & Telescope & Date & Exposure time [s]& program ID\\
\hline
\hline
WD\,J023415.51--040609.28 &X-shooter & VLT & 2019-01-11& UVB 2$\times$650, VIS 2$\times$600 & 0102.C-0351\\
WD\,J052914.32--340108.11 &X-shooter & VLT & 2019-01-11& UVB 4$\times$1250,  VIS 4$\times$1200 &  0102.C-0351\\
 &X-shooter & VLT & 2020-01-04& UVB 2$\times$1250, VIS 2$\times$1200 & 1103.D-0763\\
 &X-shooter & VLT & 2020-03-02& UVB 2$\times$1250,  VIS 2$\times$1200 & 1103.D-0763\\
WD\,J084602.47+570328.64 &FLWO&FAST&2016-01-23&1200 & 166 \\
&ISIS&WHT&2016-04-12&3$\times$1200, 3$\times$1800, 2$\times$1000 & W/2016A/26\\
&OSIRIS&GTC&2016-05-18& R2500I, $3\times600$ & GTC05-16ADDT\\
&OSIRIS&GTC&2016-09-19& R2500I, $3\times600$ & GTC1-16ITP\\
&HIRES&Keck&2017-05-22& HIRESb, $3\times900$ & N188 \\
&ISIS&WHT&2017-09-25&3$\times$1800 & SW2017b01\\
&ISIS&WHT&2019-04-14&2$\times$1000 & ITP2018/19\\

WD\,J193037.65--502816.91 &X-shooter & VLT & 2019-05-25& UVB 2$\times$581, VIS 2$\times$531 & 0102.C-0351 \\
WD\,J213350.72+242805.93  &ISIS&WHT&2019-07-05& 2$\times$900 &082-MULTIPLE-2/19A\\
WD\,J221202.88--135239.96&MagE&Baade&2016-08-24&3600 &CN2016B-76\\
 & X-shooter& VLT& 2016-09-28& UVB 2$\times$1221,  VIS 2$\times$1255 & 097.D-1029\\
  &ISIS&WHT&2017-09-24&4$\times$1800 &SW2017b01\\
 & X-shooter& VLT& 2018-05-18& UVB 2$\times$1475,  VIS 2$\times$1420& 5100.C-0407\\
 & X-shooter& VLT& 2019-07-06& UVB 2$\times$1475, VIS 2$\times$1420&5100.C-0407\\
 &ISIS&WHT&2019-08-03&3$\times$1200 & W/2019A/09\\

\hline
\end{tabular}
\end{table*}
\end{center}

\section{Stellar parameters}
The initial SEDs used to identify our targets as IR-excess candidates relied on limited optical photometry and on simplified white dwarf models which assumed a H-atmosphere in all cases. In order to correctly characterize the white dwarfs in these systems and thus quantify the strength of the thermal emission, we used model comparison to fit for the white dwarfs surface gravity, log\,$g$, effective temperature, \Teff, and H-abundance, [H/He] (for He-atmosphere white dwarfs). The newly determined stellar parameters were then used to construct more reliable SEDs for all objects in our sample (displayed in Fig.\,\ref{SEDs}). 

\begin{figure*}
	\includegraphics[width=1.9\columnwidth]{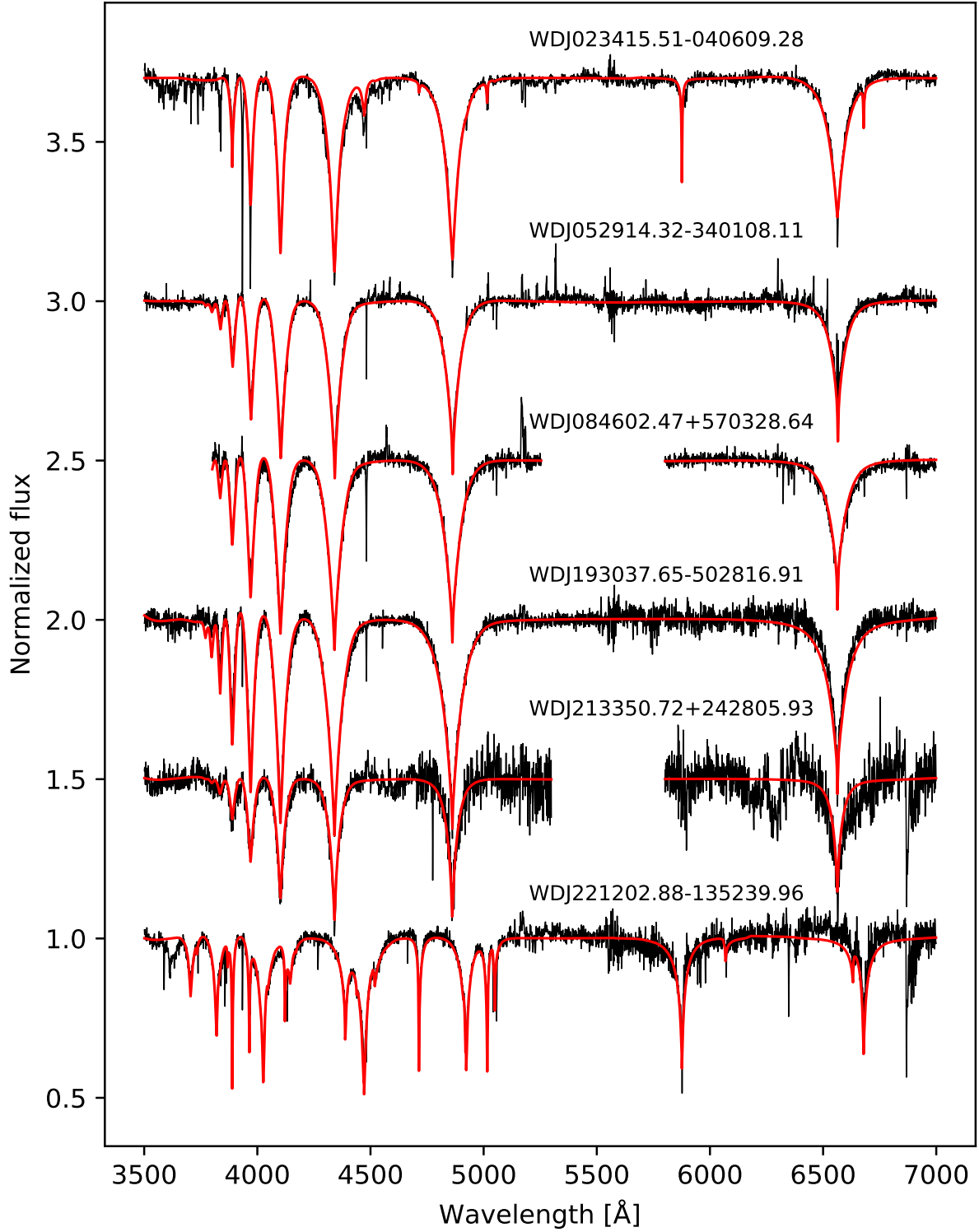}
    \caption{Best-fitting white dwarf models (red) compared with the spectroscopic observations (black) for the six objects presented in this article. All spectra and models have been continuum-normalized, and vertically offset from one. The stellar parameters corresponding to the best-fit models are reported in Table\,\ref{mega_table}}
    \label{spec_fit}
\end{figure*}

For each star we simultaneously fitted \textit{Gaia} parallaxes, continuum-normalised spectroscopy (Fig.\,\ref{spec_fit}) and optical photometry from various sources (see Table\,\ref{mega_table}) using models that match the main atmospheric composition of these white dwarfs. 

We correct our models for reddening using $E(B-V)$ values from the 3D STructuring by Inversion the Local Interstellar Medium (STILISM) reddening map \citep{lallementetal18-1}, however the colour correction terms are relatively small ($0.002-0.023$\,mag) and have only a marginal impact on the result of the fit. 
We used the model atmosphere code from \cite{koester10-1} to compute a grid of synthetic white dwarf models. We created two separate grids, one with pure H-atmospheres white dwarfs and another that considers mixed H/He-atmosphere white dwarfs. Both grids cover $9000 \leq T_{\textrm{eff}} \leq 35\,000$\,K  and $7.0 \leq \log g \leq 9.0$\,dex in steps of 200\,K and 0.2\,dex respectively, in addition to the mixed atmosphere grid spans $-5.0 \leq$ [H/He] $\leq -0.4$\,dex in steps of 0.2\,dex. We set the mixing length parameter to 0.8 for H-dominated atmospheres, and to 1.0 for He-dominated atmospheres \citep{cukanovaiteetal19-1}. 
In each iteration of our fitting procedure a reference model (obtained by interpolating between grid points) is scaled to the \textit{Gaia} parallax (allowing for the exact value to vary within the parallax uncertainty) and is compared to both the optical photometry and to the continuum-normalised  spectrum of the white dwarf. The photometric and spectroscopic data are assigned equal weight and an overall chi-squared is calculated. Subsequent reference models for the comparison are then created following the chi-squared gradient until a minimum is found. In fitting He-atmosphere white dwarfs, [H/He] is mostly determined by the strength of the hydrogen lines in the spectroscopic data. However, this abundance value is significantly correlated to the $\log$ g and so the values obtained in our fitting procedure cannot be considered independent of the astrometric and photometric data.
The stellar parameters obtained from our fits are summarised in Table\,\ref{mega_table}. 
The reported uncertainties (Table\,\ref{mega_table}) are of statistical nature only, and in the case of high signal-to-noise ratio data are likely underestimated. A detailed analysis of the underlying systematic uncertainties would require a larger sample of stars and data (see e.g. \citealt{fuchsetal17-1}), and is beyond the scope of this paper.

One caveat regarding the atmospheric parameters determined here is that we do not include metals in the computation of the atmospheric structures, and line blanketing in the ultraviolet (UV) has the potential to affect the thermal structure of the atmospheres. 

The effect of blanketing in atmospheres where hydrogen dominates the opacity has been explored by \citet{gaensickeetal12-1}, and found to be negligible. To out knowledge, a similar systematic study of line blanketing in DBZ has so far not been carried out, though brief exploratory discussions are found in \citealt{dufouretal12-1}\citealt{coutuetal19-1}). Based on the available studies, we expect the effect of line blanketing in WD\,J0234-0$4$06, WDJ\,0529-3401,  WD\,J1930-5028 and WD\,J2133+2428 to be negligible, and we may over-estimate the effective temperatures of WD\,J0846+5703,  WD\,J2212$-$1352 by up to $2-3$~per cent. A full evaluation of the line blanketing in the two latter stars will have to await far-UV spectroscopy, which is necessary to establish detailed abundances for all elements that affect the opacities of their atmospheres.


\section{IR excess modelling}
With the IR flux of the white dwarf set by the best-fit models derived in the previous section, we proceeded to characterise the IR excess due to the thermal emission of the circumstellar dust. We decided to adopt two different procedures to fit the IR excess, using both an optically-thick, geometrically- thin disc, and a single blackbody.

The common approach in modelling the SEDs of white dwarfs with dusty debris discs is to adopt the passive, geometrically-thin, optically-thick disc model from \citet{jura03-1}, see e.g. \citet{vonhippeletal07-1, kilicetal12-1, rocchettoetal15-1}. Free parameters in this model are the: temperatures at the inner and outer edges of the disc, $T_{\mathrm{in}}$, and $T_{\mathrm{out}}$, and the disc inclination, \textit{i}. Notably, a significant degeneracy exists between $T_\mathrm{in}$ and $i$, see \citet{girvenetal12-1}, in particular fig.\,5 in that paper. A caveat to bear in mind in modelling the SEDs of debris disc systems with this approach is that even if an acceptable fit of the available data is achieved, this does not imply that the model of a thin, optically-thick disc is physically correct. Hints that reality may be more complex were the observations of GD362, which \citet{juraetal07-2} modelled by adding a warp to the standard disc model. In the case of G29-38, \citet{reachetal09-1} argued that the data could be fitted with the combination of a thin, optically-thick disc plus an optically thin cloud. The subsequent detection of wide-spread variability of the emission from the dust further highlighted that the standard model is likely not a fully adequate description of the real distribution of the debris \citep{xuetal14-2,xuetal18-1, swanetal19-1,swanetal20-1,rogersetal20-1}. 

The cool dust at the outer edge of a debris disc displays significant emission only at wavelengths longer than those covered by the \textit{WISE} $W1$ and $W2$ bands. Consequently $T_{\mathrm{out}}$ is unconstrained in our fits to the available data. In order to obtain a statistically significant solution, we decided to fix $T_{\mathrm{out}}$ to the typical value of 600\,K \citep{juraetal07-1}.
We find that the SEDs of the newly discovered systems can be modelled adequately with a the sum of the white dwarf plus the standard disc model within a range of physically reasonable parameters (Fig.\,\ref{SEDs}), with one exception: WDJ\,0846+5703 provides the first example where an optically-thick, geometrically-thin disc completely fails to reproduce the detected IR excess. This system has the highest fractional IR brightness ($\tau=5.8$\,per cent) of any known debris disc white dwarf, exceeding the maximum value expected for a face on disc by 40\,per cent (see fig.\,4 in \citealt{rocchettoetal15-1}). 

Given that there is at least one system where the standard model fails, and that it is now clear that combining infrared data obtained at different epochs is subject to systematic uncertainties because of the intrinsic variability, we decided to also fit the observed SEDs with the sum of a white dwarf and a single blackbody, both located at the distance of the source measured by \textit{Gaia}. The free parameters in this case are the temperature of the blackbody, $T_{\textrm{BB}}$, and its area, $A_{\textrm{BB}}$. This prescription has the advantage that it is independent of any underlying physical assumption, enables a quantitative comparison with other studies of debris disc systems, and also provides a simple metric to exclude the possibility that the detected IR excess is caused by a brown dwarf companion. With the possible exception of WD\,J0234$-$0406, all $A_{\textrm{BB}}$ values exceed that of a typical brown dwarf by one order of magnitude \citep{sorahanaetal13-1}. We use the ratio of the integrated blackbody and white dwarf fluxes to compute the fractional IR brightness, $\tau$, of the debris disc (Table\,\ref{mega_table}). With the aforementioned exception of WD\,J0846+5703, we find the $T_{\textrm{BB}}$ and $\tau$ values to be broadly consistent with those of previously known white dwarfs with dusty debris disc systems \citep{rocchettoetal15-1,dennihyetal17-1,dennihyetal20-2}.

\afterpage{
\clearpage
\begin{landscape}
\begin{table}
\caption{\label{mega_table} Stellar parameters, photometry used for SED modelling, parameters of blackbody fit to the IR flux, and properties of the \ion{Ca}{II} emission lines for all stars in our sample.}
\begin{tabular}{lD{?}{\,\pm\,}{5.3}D{?}{\,\pm\,}{5.3}D{?}{\,\pm\,}{5.3}D{?}{\,\pm\,}{5.3}D{?}{\,\pm\,}{5.3}D{?}{\,\pm\,}{5.3}}
\hline
 & \multicolumn{1}{c}{WD\,J0234--0406} &  \multicolumn{1}{c}{WD\,J0529--3401} &  \multicolumn{1}{c}{WD\,J0846+5703} & 
 \multicolumn{1}{c}{WD\,J1930--5028} & \multicolumn{1}{c}{WD\,J2133+2428} & \multicolumn{1}{c}{WD\,J2212--1352}\\
\hline
type & \multicolumn{1}{c}{DABZ} & \multicolumn{1}{c}{DAZ} & \multicolumn{1}{c}{DAZ} & \multicolumn{1}{c}{DAZ} & \multicolumn{1}{c}{DAZ} & \multicolumn{1}{c}{DBZ} \\
Opt. phot. & \multicolumn{1}{c}{SDSS $u,g,r,i,z$} & \multicolumn{1}{c}{SkyMapper $u,v,g,r,i$} & \multicolumn{1}{c}{SDSS $u,g,r,i,z$} &  \multicolumn{1}{c}{SkyMapper $u,v,g,r,i,z$} & \multicolumn{1}{c}{SDSS $u,g,r,i,z$} & \multicolumn{1}{c}{ATLAS $u,g,r,i,z$}\\
 & \multicolumn{1}{c}{PanSTARRS $g,r,i,z,Y$} & &\multicolumn{1}{c}{PanSTARRS $g,r,i,z,Y$} & &
 \multicolumn{1}{c}{PanSTARRS $g,r,i,z,Y$} &
 \multicolumn{1}{c}{PanSTARRS $g,r,i,z,Y$}\\
\Teff [K]& 13454?200 & 23197?300 & 17803?500 & 13306?500 & 29282?400 &24969?400\\
log $g$ & 8.01?0.03 &8.01?0.04 & 8.02?0.05 & 7.90?0.05 & 7.85?0.03 & 7.94?0.03 \\
$\mathrm{[H/He]}$ & \multicolumn{1}{c}{--1.99}
&\multicolumn{1}{c}{-} & \multicolumn{1}{c}{-} & 
\multicolumn{1}{c}{-} & \multicolumn{1}{c}{-} & \multicolumn{1}{c}{-}\\
Mass [$M_{\odot}$]& \multicolumn{1}{c}{0.59} & \multicolumn{1}{c}{0.64} &\multicolumn{1}{c}{0.63} &\multicolumn{1}{c}{0.55} &\multicolumn{1}{c}{0.57} &\multicolumn{1}{c}{0.58}\\
Metal abs. & \multicolumn{1}{c}{Mg, O, Ca, Al, Ti, Fe} &\multicolumn{1}{c}{Mg, Si, Ca} & \multicolumn{1}{c}{Si, Mg} & 
\multicolumn{1}{c}{Mg, Ca} & \multicolumn{1}{c}{-} & \multicolumn{1}{c}{Mg, Si, C, Ca, Al, Fe}\\
Metal emi. & \multicolumn{1}{c}{Ca, Mg/Fe} & \multicolumn{1}{c}{H, Ca, Mg, O, Fe} & \multicolumn{1}{c}{Ca, Mg, Fe} & \multicolumn{1}{c}{Ca, Mg, Fe}& \multicolumn{1}{c}{ Ca, O } & \multicolumn{1}{c}{Ca, Mg, Fe}\\
\hline
IR phot. & \multicolumn{1}{c}{VHS {\it J, H, Ks}} & \multicolumn{1}{c}{VHS {\it H, Ks}} & \multicolumn{1}{c}{2MASS {\it J, H, Ks}} & \multicolumn{1}{c}{VHS {\it H, Ks}} &
\multicolumn{1}{c}{UHS {\it J}} &
\multicolumn{1}{c}{VHS {\it J, H, Ks}}\\
& \multicolumn{1}{c}{WISE {\it W1, W2}} & \multicolumn{1}{c}{WISE {\it W1, W2}} & \multicolumn{1}{c}{WISE {\it W1, W2}} & \multicolumn{1}{c}{WISE {\it W1, W2}} & \multicolumn{1}{c}{WISE {\it W1, W2}} & \multicolumn{1}{c}{WISE {\it W1, W2}}\\
$T_{\mathrm{bb}}$ [K]& \multicolumn{1}{c}{1226} & \multicolumn{1}{c}{1010} & \multicolumn{1}{c}{1165} & \multicolumn{1}{c}{949} & \multicolumn{1}{c}{1442} & \multicolumn{1}{c}{1266}\\
$A_{\mathrm{\,bb}}$ [$10^{21}$cm$^2$]$^{1}$& \multicolumn{1}{c}{0.15} & \multicolumn{1}{c}{9.16} & \multicolumn{1}{c}{7.85} & \multicolumn{1}{c}{0.80} & \multicolumn{1}{c}{3.85} & \multicolumn{1}{c}{3.63}\\
$\tau [\%]$ & \multicolumn{1}{c}{0.39} & \multicolumn{1}{c}{1.27} & \multicolumn{1}{c}{5.82} & \multicolumn{1}{c}{0.75} & \multicolumn{1}{c}{0.68} & \multicolumn{1}{c}{0.37}\\
\hline  \\ [-2.5 ex]
Ca 8498 EW [\AA] & 1.45?0.10^{2,3} & 1.58?0.03^{5}& 9.53?0.08^{6}&8.40?0.16^{2}& 9.26?0.45&6.71?0.08^{7}\\
Ca 8542 EW [\AA] & 1.45?0.10^{2,3}& 1.69?0.04^{5}& 12.50?0.08^{6} & 8.40?0.16^{2} & 8.42?0.93& 7.63?0.07^{7}\\
Ca 8662 EW [\AA] & 0.94?0.09^{3}& 1.37?0.05^{5}& 11.89?0.09^{6}&4.08?0.14& 8.82?0.51&7.20?0.08^{7}\\
FWZI [km\,s$^{-1}$] & 2300?100^{4} & 301?6^{5} & 372?6^{6} & 1900?100^{4} & 1260?30 & 1000?20^{7} \\
\hline
\hline
\vspace{-3mm}\\
\multicolumn{6}{l}{$^{1}$ For reference, a circular disc of radius 1\,Rsun has area of 1.52 $10^{21}$cm$^2$}\\
\multicolumn{6}{l}{$^{2}$ blended emission lines measured as one.}\\ 
\multicolumn{6}{l}{$^{3}$ Strongly contaminated by photospheric absorption feature(s). EW quoted is a lower limit.}\\
\multicolumn{5}{l}{$^{4}$ FWZI determined from 8662\, \AA\ feature only.}\\ 
\multicolumn{5}{l}{$^{5}$ 2019-01-11 observation. Maximum recorded EW value.}\\ 
\multicolumn{5}{l}{$^{6}$ 2016-04-12 observation. Maximum recorded EW value.}\\ 
\multicolumn{5}{l}{$^{7}$ 2016-09-28 observation. Maximum recorded EW value.}\\ 
\end{tabular}
\end{table}
\end{landscape}
\clearpage}

\section{Notes on individual systems}

Here we present an overview of the most distinctive observational properties of the individual debris disc systems  in our sample. It is important to consider that the disc $i$, composition and geometry (which are unknown), as well as the white dwarf \Teff\ and mass, all affect the overall appearance of the gas emission features. Furthermore, degeneracy between these parameters is hard to break without additional information. In the following sections we attempt to provide the most comprehensive physical description of our systems, given the data currently available.

\subsection{\texorpdfstring{WD\,J023415.51--040609.28}{WD J023415.51-040609.28}}

Despite its spectral appearance visually dominated by broad Balmer absorption lines (Fig.\,\ref{spec_fit}), the atmosphere of WD\,J0234--0406 is actually He-dominated, making this star the first white dwarf of spectral type DABZ found to host a gaseous debris disc. 
With $\mathrm{[H/He]}=-1.99$, WD\,J0234--0406 stands out as extremely H-rich among He-dominated white dwarfs with similar effective temperatures (see fig.\,5 of \citealt{rolland18}).
The input physics and methods for the envelope calculations for helium/carbon diffusion equilibrium in \citet{koesteretal20-1} can also be applied to hydrogen/helium equilibrium, and so be used to calculate the mass of H in the convection zone of WD\,J0234--0406.
Following this prescription, without accounting for convective overshoot \citep{cunninghametal18-1}, the photospheric abundance translates to a total of $\simeq 2\times 10^{24}$\,g of H in the convection zone \footnote{With the inclusion of overshoot in the calculation the H mass estimate can be as high as $\simeq 5.4\times 10^{24}$\,g. However, since the effect of overshoot is not included in the majority of other publications a direct comparison of this value with other systems studied is not possible at the moment}. Such large amounts of H have so far only been detected in a handful of debris-accreting white dwarfs (fig.\,5 of \citealt{gentilefusilloetal17-1}).

WD\,J0234--0406 is also the white dwarf with the strongest metal absorption features in our sample and we were able to identify pollution from Mg, O, Ca, Al, Ti, and Fe. The presence of large amounts of H in a growing number of cool, metal-polluted He-atmosphere white dwarfs has been linked with accretion of water-bearing debris (e.g. GD\,362, GD\,61, SDSS\,J1242+5226, GD\,16, GD\,17, WD\,J204713.76--125908.9 \citealt{jura+xu10-1, farihietal11-1,raddietal15-1, koesteretal05-1,  gentilefusilloetal17-1, hoskinetal20-1}), a theory which can be corroborated if an O-excess is determined from the abundance of the accreted metals (e.g. \citealt{kleinetal10-1, farihietal13-1}). The amount of H in the convection zone of WD\,J0234--0406 is too high to be explained solely by a primordial origin \citep{rolland18} leading to the tempting speculation that this white dwarf, too, is accreting  water (or may have accreted it in the past since H never diffuses out of a white dwarf photosphere). 

\begin{figure}
\centering
	\includegraphics[width=0.88\columnwidth]{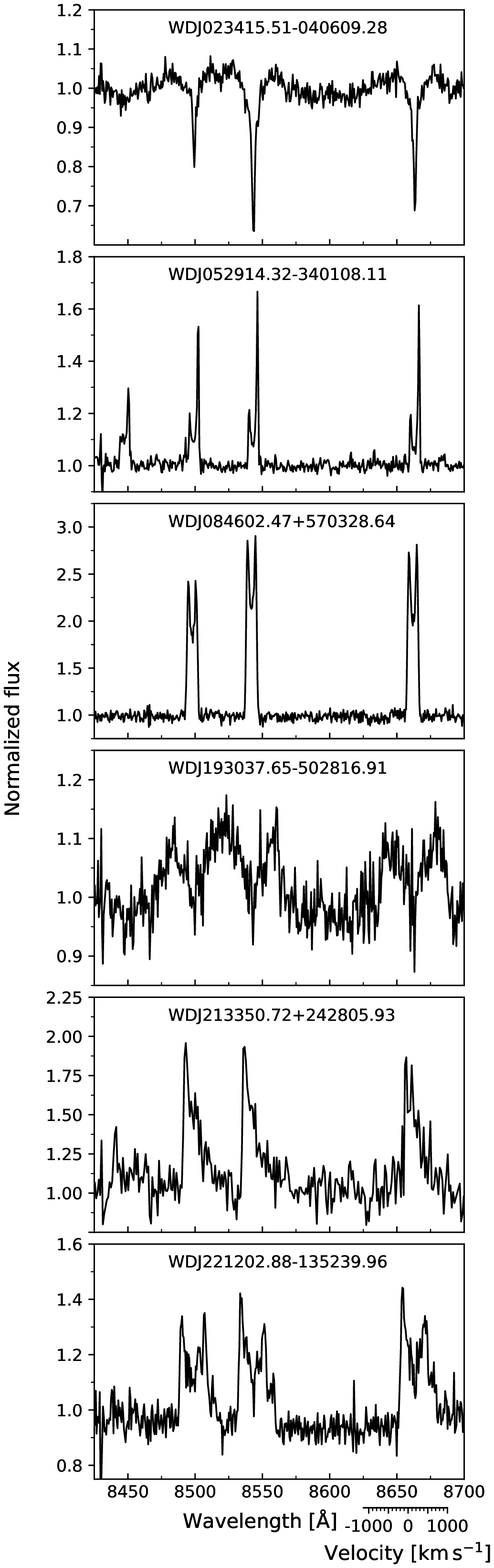}
    \caption{8448\,\AA\ \ion{O}{i} (if present) and \ion{Ca}{II} triplet emission profiles of the six newly identified gaseous debris disc hosts.}
    \label{CaII}
\end{figure}

The only emission features from the gas disc around WD\,J0234--0406 are the \ion{Ca}{II} triplet (Fig.\,\ref{CaII}) and the indistinguishable blend of \ion{Fe}{II}/\ion{Mg}{I} lines at $\simeq 5175$\,\AA. 
Due to the very large width of the \ion{Ca}{II} emission lines, we quote a combined equivalent width (EW) for the severely blended 8498\,\AA\ and 8542\,\AA\ lines. Additionally, the deep and broad \ion{Ca}{II} photospheric absorption lines in WD\,J0234--0406 inevitably affect our EW estimates which should, therefore, be considered as lower limits.

The full widths at zero intensity (FWZIs) of the \ion{Ca}{II} triplet profiles are the largest of any known gaseous debris disc at 2300\,$\pm$\,100\,km\,s$^{-1}$. The emission profiles appear relatively symmetric which could suggest a circular disc, which combined with the mass of the white dwarf of $M_{\textrm{WD}}$\,=\,0.59\,\Msun, would imply an inner edge of the disc of $r_{\textrm{in}}$\,$\simeq$\,0.085\,$\sin^2 i$\,\Rsun. Even for an edge-on configuration with $i$\,=\,90\,$^{\textrm{o}}$, this is significantly closer to the white dwarf than any other gaseous planetary material observed in emission, and may only be possible due to the relatively cooler \Teff\ of the host star.
A lower \Teff\ means that the photo-ionised gas responsible for the visible emission could be closer to the white dwarf compared to other systems. Close-in gas would orbit at higher velocity, leading to larger Doppler broadening of the lines for a given value of $i$. While we cannot constrain the inclination from the profile shapes due to the photospheric contamination of the emission profiles, we expect this system to be in a close to edge-on configuration.

\subsection{\texorpdfstring{WD\,J052914.32--340108.11}{WD J052914.32-340108.11}}

\begin{figure*}
	\includegraphics[width=1.9\columnwidth]{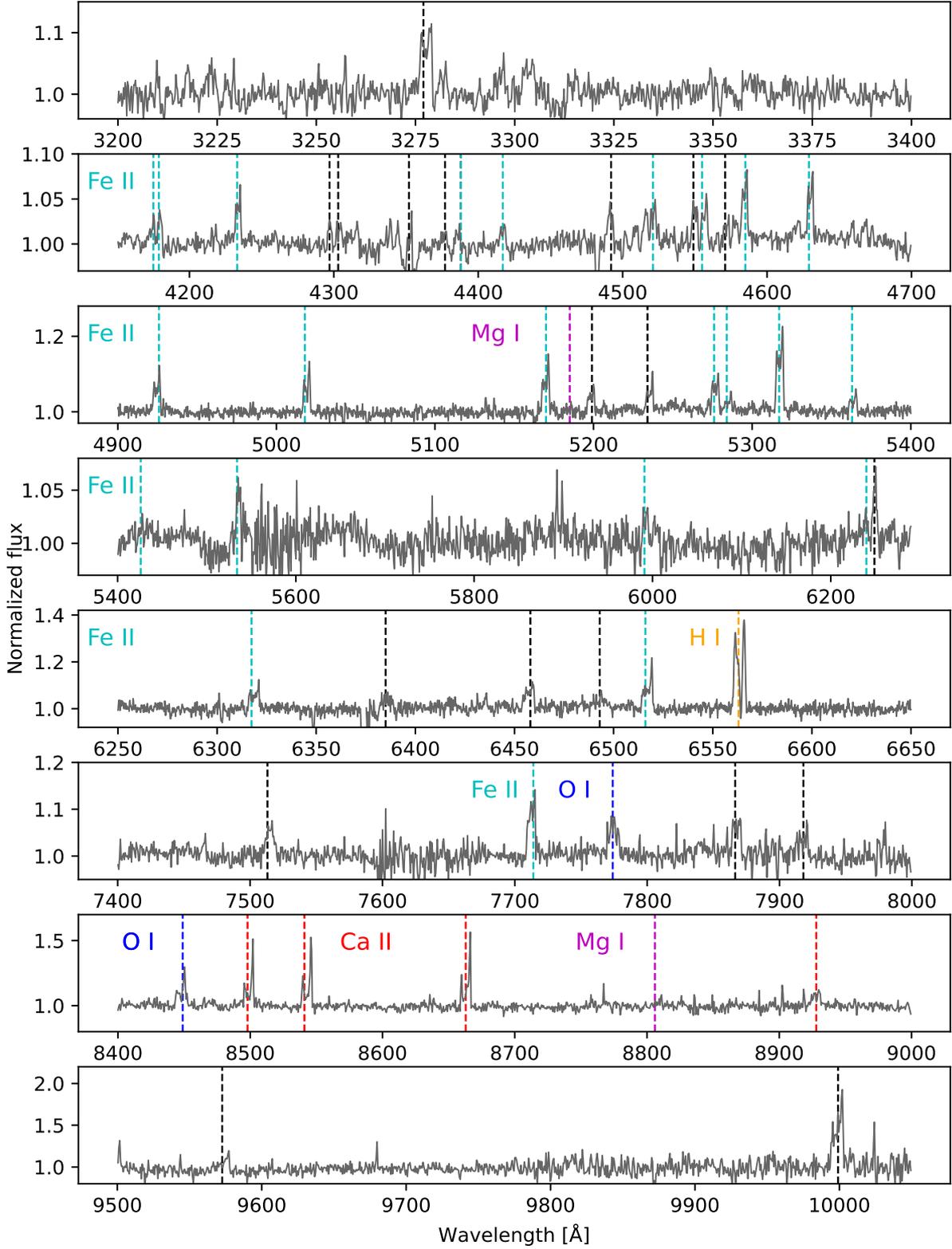}
    \caption{All emission features identified in the spectrum of WD\,J0529--3401 reported in Table\,\ref{all}. Features marked by black vertical lines have not been yet associated with an element. The spectrum has been continuum normalised, and the photospheric Balmer and Paschen lines have been flattened for clarity. We note that the yet unidentified emission feature at 9999\,\AA\ is stronger than the \ion{Ca}{II} triplet. We speculate this may be an Fe fluorescence line, see Sect.\,6.2}
    \label{WDJ0529_emi}
\end{figure*}

WD\,J0529--3401 has the second brightest debris disc of our sample in terms of $\tau$, but what truly makes this system stand out is its array of emission features, not matched in number and diversity by any other known planetary debris host. We identified emission lines of H, Mg, Ca, Fe and O; though there are also a number of yet unidentified features (Fig.\,\ref{WDJ0529_emi}).
The unique spectral appearance of this system is most likely due to a combination of factors. The high \Teff\  of the white dwarf may be partially responsible; indeed the recently discovered WD\,J210034.65+212256.89 \citep{dennihyetal20-2, melisetal20-1} has \Teff $\simeq 26\,000$\,K and also displays several emission lines from elements other than Ca. 
However, other similarly hot and even hotter debris disc systems (e.g. WD\,J2133+2428 and WD\,J2212--1352) do not share the same variety of emission features (see Table\,\ref{full_sample}). Other factors like a peculiar debris composition (possibly richer in Fe compared to those about other emission line systems e.g \citealt{hollandsetal18-1}) and/or the disc geometry and distance from the star should also play a significant role in determining which emission lines are visible.

A particularly noteworthy emission feature in WD\,J0529--3401 is that of H$\alpha$, which is not observed in any other white dwarf with a gaseous debris disc. The only related system is WD\,J091405.30+191412.25, a white dwarf hosting a gaseous disc that exhibits emission lines of only H, O, and S in the optical wavelength range \citep{gaensickeetal19-1}. No rock-forming elements (Mg, Ca, Fe) are detected in emission from that disc, or in absorption in the white dwarf photosphere, leading to the interpretation that this white dwarf is evaporating a giant planet, and accreting the purely volatile atmospheric material.
In contrast, WD\,J0529--3401 displays circumstellar dust in the form of a strong IR excess, as well as rock-forming elements in its photosphere, clearly indicating that this white dwarf is accreting rocky debris. However, the H$\alpha$ emission unequivocally proves that H is also present in the debris together with O and other rock forming elements, strongly suggesting the presence of water in the planetesimal currently being accreted. As mentioned above, a number of white dwarfs exhibit an O-excess in the abundances of the accreted debris, and are suspected to have disrupted water-rich planetesimals. The detection of an O-excess in WD\,J0529--3401 would corroborate that interpretation, but will require a detailed abundance analysis. 

Whereas the majority of the over 50 distinct emission lines visible in the X-shooter spectrum of WD\,J0529--3401 are associated with \ion{Fe}{II}, there remain a number of features which we were unable to identify (Fig.\,\ref{WDJ0529_emi}, Table\,\ref{all}). In particular the emission line at $\simeq 9999$\,\AA, is the strongest one in the entire spectrum exceeding even the \ion{Ca}{II} lines in EW (Table\,\ref{all}). Active Galactic Nuclei (AGN) \citep{landtetal08-1}, luminous blue variables \citep{ritchieetal09-1}, and classical novae \citep{andrillat+houziaux94-1} often display an \ion{Fe}{II} line near that wavelength, which is a Ly$\alpha$-pumped fluorescence line \citep{sigut+pradhan03-1}. The emission line detected near 3277\,\AA\ may have the same origin. Given that we detect H$\alpha$ emission in WD\,J0529$-$3401, it is feasible that also Ly$\alpha$ is in emission. There are also numerous additional potential emission features near the noise level that will require higher-quality data for an unambiguous detection.

\begin{figure}
	\includegraphics[width=\columnwidth]{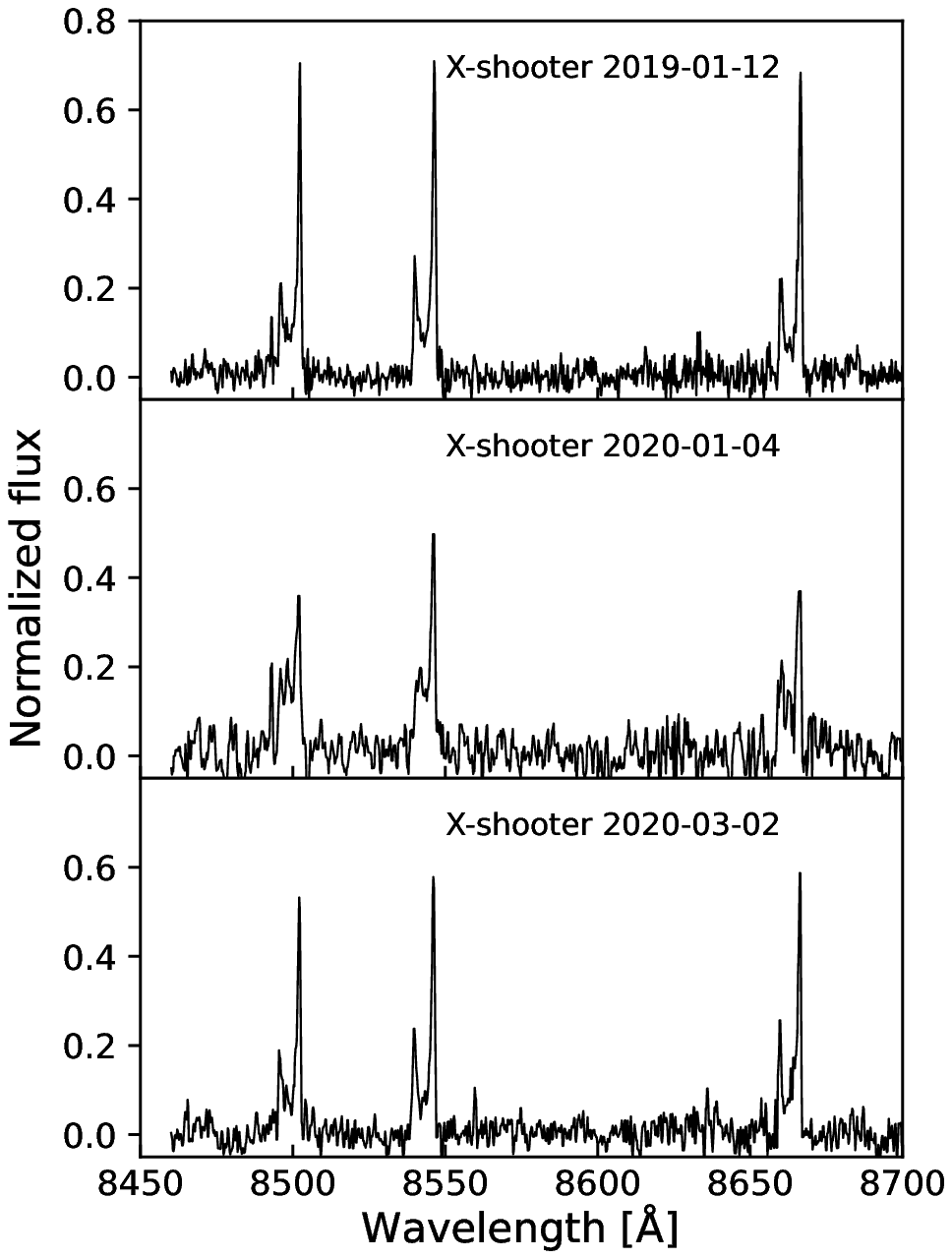}
\includegraphics[width=.98\columnwidth]{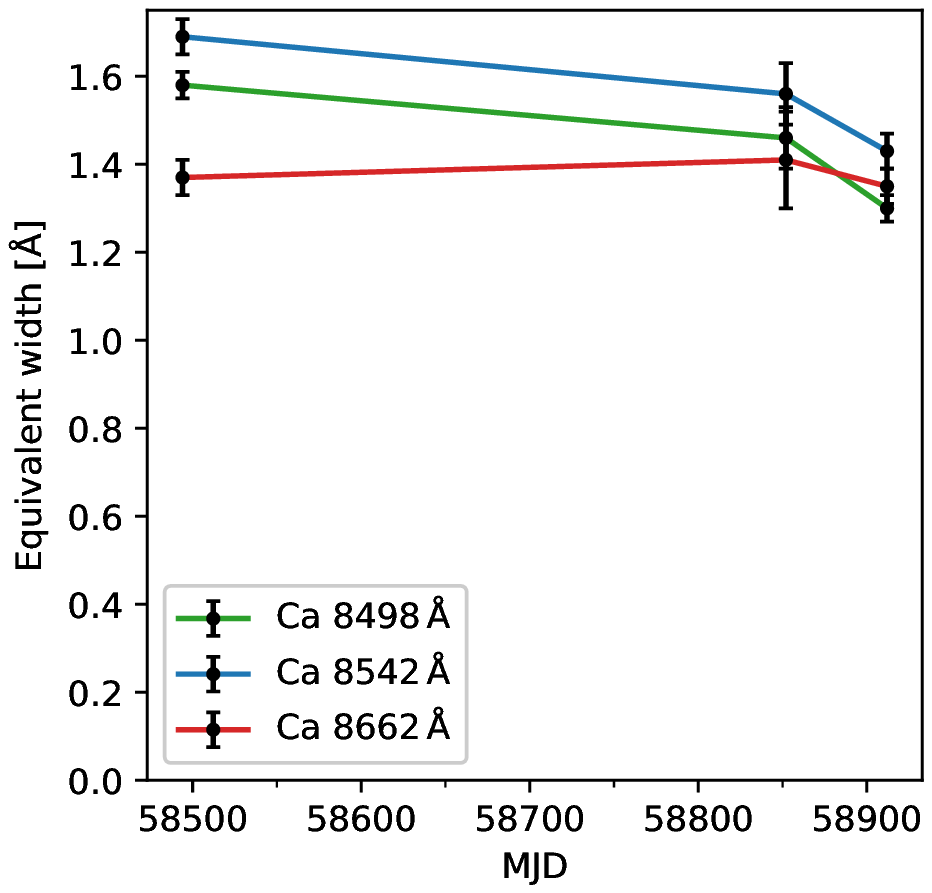}
    \caption{\textit{Top panel:} Different epochs of spectroscopic observations of the \ion{Ca}{II} emission lines in  WD\,J0529--3401.
    \textit{Bottom panel:} Change in the equivalent width of the \ion{Ca}{II} triplet emission lines of WD\,J0529--3401.}
    \label{WDJ0529_epoch}
\end{figure}

To date we have acquired three epochs of spectroscopic data for WD\,J0529--3401 (Fig\,\ref{WDJ0529_epoch}). While these spectra only span $\simeq$1.14\,yr, they reveal a potential decrease in the overall strength of the emission lines (Fig.\,\ref{WDJ0529_epoch}), along with a possible change in the morphology of the profiles, with the two more recent spectra showing substantially reduced blue- and red-shifted emission peaks compared to the observation. Further monitoring of this system will be required to confirm morphological variability, which might occur on a similarly short time-scale to that of the gaseous debris disc around HE\,1349--2305 \citep{dennihyetal18-1}.

\subsection{\texorpdfstring{WD\,J084602.47+570328.64}{WD J084602.47+570328.64}}

While this system was identified by \cite{melisetal20-1} as a gaseous debris disc host, we independently determined the presence of a gas disc a this system and present our results and discussion below. The initial FAST spectrum of WD\,J0846+5703 did not cover the \ion{Ca}{II} triplet, but the presence of a gas disc was revealed by the blended \ion{Mg}{I}/\ion{Fe}{II} emission feature at $\simeq$5175\,\AA.
WD\,J0846+5703 stands out as the brightest debris disc about a white dwarf (in terms of fractional IR luminosity, $\tau=5.82$\,per\,cent), exceeding the brightness of the disc around both GD\,362 \citep{becklinetal05-1} and the recently discovered WD\,J061131.70$-$693102.15 \citep{dennihyetal20-2}. 
A fit adopting the standard optically-thick, geometrically-thin disc model \citep{jura03-1} fails to reproduce the observed SED.
 Given both the distance to, and the stellar parameters of, the white dwarf, even the brightest possible disc configuration ($i = 0^{\textrm{o}}$) underestimates the fractional IR luminosity by $\simeq$40\,per\,cent. A simplistic single-temperature blackbody model provides a statistically acceptable fit to the IR flux of WD\,J0846+5703 (Fig.\,\ref{SEDs}), however, the physical interpretation of the derived parameters remains uncertain.

A physically meaningful model of the IR emission of WD\,J0846+5703, may need to invoke multiple dusty components (as proposed for GD\,362 \citealt{juraetal07-2}), or a configuration different from a geometrically-thin, optically-thick disc. In fact, recent \textit{Spitzer} observations suggest that a variable, optically-thin component is most likely present at most debris discs, with gaseous disc hosts being the most variable \citep{swanetal20-1}. 

Whereas our estimate for the total IR flux of WD\,J0846+5703 relies on \textit{WISE} photometry and contamination from background sources could impact both the overall SED shape and the value of our $\tau$ \citep{dennihyetal20-1}, near-IR imaging obtained with the Long-slit Intermediate Resolution Infrared Spectrograph on the WHT (Fig.\,\ref{LIRIS}) and mid-IR \textit{Spitzer} observations \citep{swanetal20-1} confirm that the extremely large IR fluxes detected by \textit{WISE} are associated with WD\,J0846+5703. 

\begin{figure}
    \centering
    \includegraphics[width=\columnwidth]{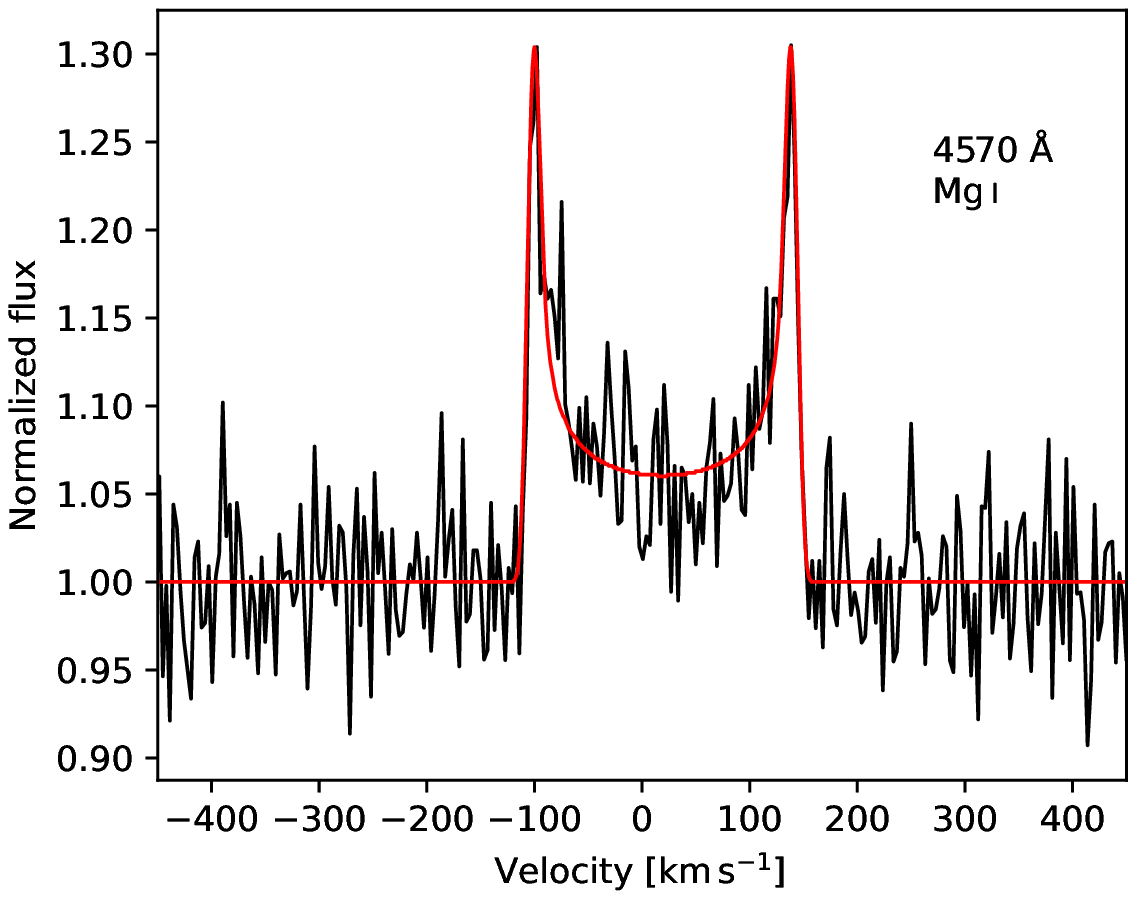}
    \caption{HIRES observations of the \ion{Mg}{i}~4571\,\AA\ emission line in WD\,J0846+5703 (black). Superimposed is an example of a model fit of a thin ring with an inner and outer radius of 1.0\,\Rsun\ and 1.1\,\Rsun\ seen at an inclination of $21^\circ$. }
    \label{f:wdj0846_mgi}
\end{figure}

The profile of the \ion{Ca}{II} emission lines of WD\,J0846+5703 is particularly narrow (Fig.\,\ref{CaII}), with $\mathrm{FWZI}=372\pm6\,\mathrm{km\,s^{-1}}$.
The higher resolution of the HIRES spectrum reveals additional emission lines which display extremely narrow blue and red-shifted peaks (Fig.\,\ref{f:wdj0846_mgi}), suggesting that the emission arises over a limited range in radii within the disc. We chose to model the \ion{Mg}{i}~4571\,\AA\ line, which is in a region of the spectrum where the S/N of the HIRES data peaks, and which is not blended with any other emission or absorption line. We used the analytical expressions for the Abel transform of \citet{smak81-1} to compute the intensity of the gaseous disc, where we adopted a power-law index of zero for the radial intensity distribution, and varied the inner and outer radius of the emitting region to match the morphology of the \ion{Mg}{i} profile. We found that the observed profile shape is best reproduced by a very narrow ring, where the outer and inner radii are $R_\mathrm{out}=1.1\times R_\mathrm{in}$. The value of $R_\mathrm{in}$ is degenerate with the inclination, i.e. $R_\mathrm{in}\propto(\sin i)^{2}$. The symmetric shape of the observed \ion{Mg}{i} line is consistent with a circular distribution of the emitting gas. However, we cannot rule out the possibility that we are viewing an eccentric disc along the semi-major axis, which would also present a symmetric distribution.

Interestingly WD\,J0846+5703 displays no emission of \ion{O}{I}~7774/8446\,\AA, which is detected in gaseous debris discs at white dwarfs of similar \Teff, implying that the absence of this feature may not relate to the \Teff\  of the star. One might speculate about a particular O-poor composition of the planetary debris \citep{harrisonetal18-1,doyleetal20-1}. Yet, the detection of both Mg and Ca emission lines implies the presence of O within the gas, as both elements are bound to O when in solid mineral form. One possible explanation for the lack of \ion{O}{i} lines may be related to the fact that these arise from high-excitation transitions~--~if the gas is located sufficiently far from the white dwarf, these transitions may not occur. This hypothesis is supported by the Doppler maps of SDSS\,J1228+1040, which show that the emission of the (low-excitation) \ion{Ca}{ii}~8600\,\AA\ triplet originates mainly at distances $\simeq0.6-1.2\,\Rsun$, whereas the \ion{O}{i}~7774\,\AA\ arises from closer to the white dwarf, $\simeq0.2-0.6\,\Rsun$ (fig.\,7.1 in \citealt{manser18-1}). Exploring this scenario, we fixed $R_\mathrm{in}=1.0\,\Rsun$, which then implies an inclination of $\simeq21^{\circ}$. Modelling higher S/N observations of the different emission lines both in terms of their strength and morphology has the potential to constrain both the location of the gas, and its physical conditions \citep{gaensickeetal19-1}.

\begin{figure}
	\includegraphics[width=0.95\columnwidth]{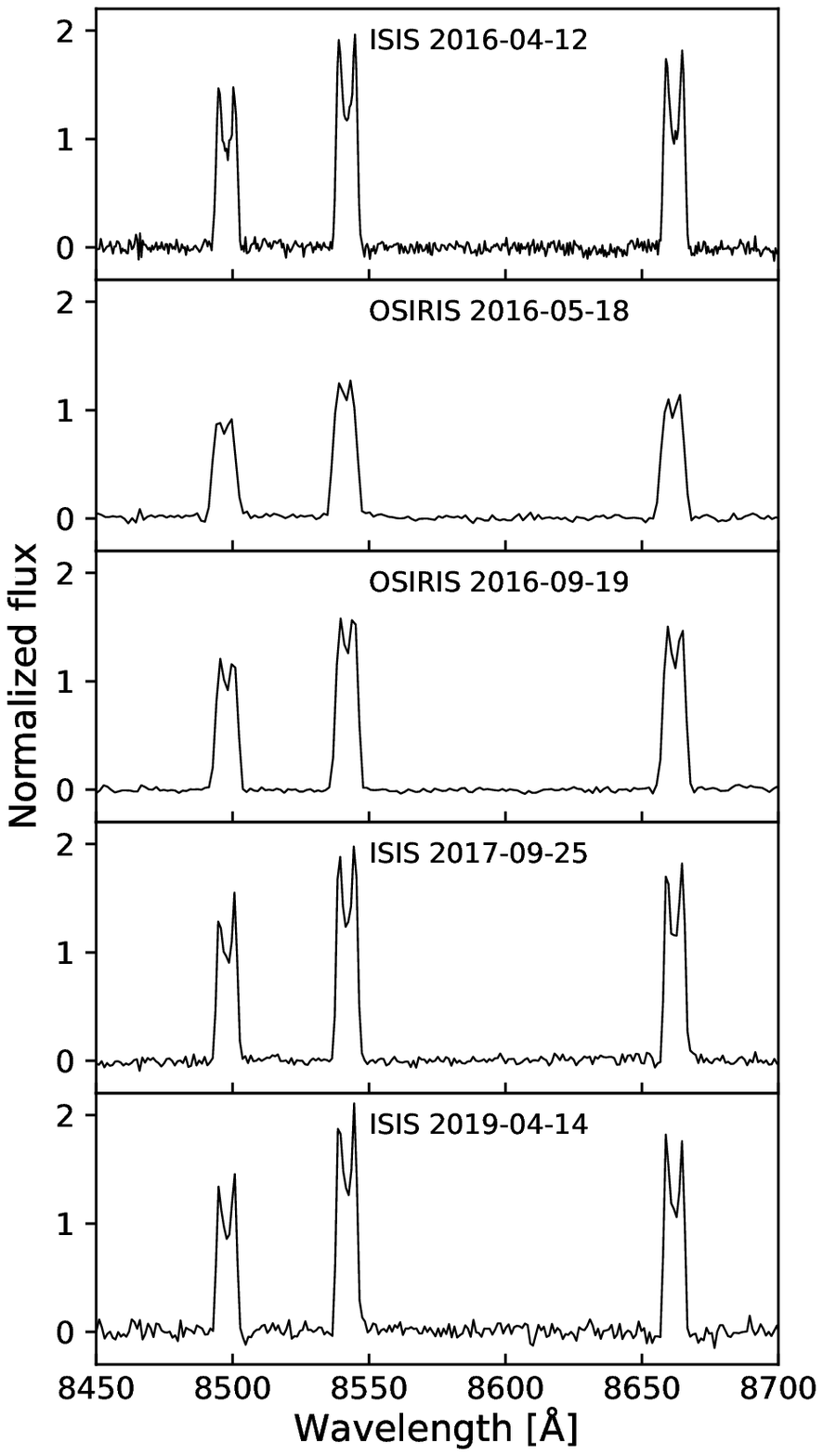}
	\includegraphics[width=0.95\columnwidth]{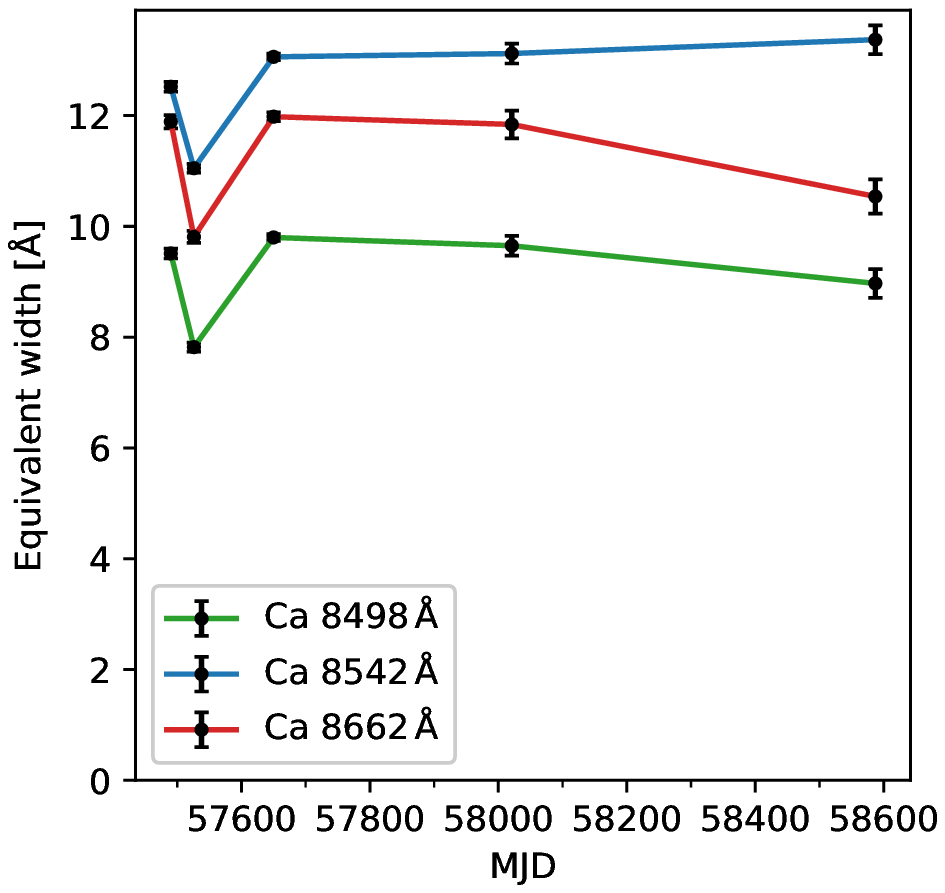}

    \caption{\textit{Top panel}: Different epochs of spectroscopic observations of the \ion{Ca}{II} emission lines in WD\,J0846+57032. The less pronounced double peaked shape of the lines in OSIRIS spectra is due to the lower resolution compared to ISIS.
    \textit{Bottom panel}: Change in the equivalent width of the \ion{Ca}{II} triplet emission lines of WD\,J0846+5703.}
    \label{WDJ0846_epoch}
\end{figure}

Since its discovery in 2016 we have obtained four additional epochs of spectroscopy for WD\,J0846+5703 covering the \ion{Ca}{II} triplet (Fig.\,\ref{WDJ0846_epoch}). Over the space of three years the \ion{Ca}{II} emission lines in this system did not undergo any significant change in shape maintaining a mostly symmetric profile, while the overall strength fluctuated by $\simeq15$\,per\,cent (Fig.\,\ref{WDJ0846_epoch}), consistent with monthly to yearly changes in the EW reported for SDSS\,J1228+1040 \citep{manseretal16-1,manseretal19-1}. This limited variability could indicate that the gas in this system follows a near-circular orbit, with a weak observational signature in the lines shape. While the low-resolution spectra of the narrow Ca\,{\textsc{ii}} triplet emission of WD\,J0846+5703 make it hard to constrain any morphological variability, \cite{melisetal20-1} identify the same Mg\,{\textrm{i}} 4571\,\AA\ feature we present in Fig.\,\ref{f:wdj0846_mgi} in their 2019 December and 2020 June HIRES spectra. The profile is similarly symmetric, suggesting no significant variability has taken place over a $\simeq$\,3\,yr time-frame. This is also supportive of a circular ring of material, however we cannot currently rule out a slowly precessing eccentric ring viewed along the semi-major axis.
Confirming a circular orbit would allow us to constrain the location of the gas without the need of long timeline repeated observations making WD\,J0846+5703 a particularly valuable target to study the structure of gaseous debris discs.

\subsection{\texorpdfstring{WD\,J193037.65--502816.91}{WD J193037.65-502816.91}}

Similar to WD\,J0234--0406, the line profiles of the \ion{Ca}{II} emission in WD\,J1930--5028 are very broad ($\mathrm{FWZI}=1900\pm100$\,km\,s$^{-1}$), to the point that the 8498\,\AA\ and 8542\,\AA\ lines are blended together making EW measurements of the individual profiles impossible. We therefore quote the EW of these two profiles as a single value in Table\,\ref{mega_table}. The combined EW is roughly a factor two larger than the 8662\,\AA\ component, consistent with a visual inspection of the profiles which appear approximately equal height outside of the blended region. 
Using $M_{\textrm{WD}}=0.55$\,\Msun\ we estimate the location of the inner edge of disc about WD\,J1930--5028 to be at $r_{\textrm{in}}\simeq\,0.12\,\sin^2 i$\,\Rsun. Similar to WDJ\,0234--5028, this inner edge would suggest a configuration close to edge-on, again with the low \Teff\ of the system likely being a contributing factor. Furthermore, the zero-velocity center of the emission lines is relatively deep, particularly compared to WD\,J0529--3401 and WD\,J0846+5703, suggesting a significant level of self-absorption, which should be higher at lower inclinations \citep{horne+marsh86-1}.

\subsection{\texorpdfstring{WD\,J213350.72+242805.93}{WD J213350.72+242805.93}}

With a $\Teff=29\,282$\,K, WD\,J2133+2428 is, by a wide margin, the hottest known white dwarf with a debris disc \citep{xuetal15-1, xuetal19-1}, even hotter than WD\,J091405.30+191412.25, the white dwarf which was recently reported to be accreting gas from an evaporating  giant planet (\Teff$=27\,750$\,K  \citealt{gaensickeetal19-1}). This system pushes the boundaries of the known parameter space spanned by debris discs hosts, effectively enabling us to investigate how the properties of the observed emission lines vary with white dwarf \Teff\ between $\simeq12\,000$--$30\,000$\,K, and to extend dynamical theories about planetary remnants around such young white dwarfs  \citep{verasetal20-1}. 
As the hottest debris disc host known, WD\,J2133+2428 is also the youngest, with a cooling age of $\simeq$\,10\,Myr\footnote{http://www.astro.umontreal.ca/$\sim$bergeron/CoolingModels} \citep{bedardetal20-1}. We do not expect there to be many systems significantly hotter than WD\,J2133+2428, as above $\simeq$\,32\,000\,K irradiation from the white dwarf would sublimate even micron-sized dust grains at a distance greater than the tidal disruption radius for a rocky body \citep{vonhippeletal07-1}. 

All previously known \ion{Ca}{II} emitters display evidence of accretion from rocky planetary debris, but WD\,J2133+2428 is still sufficiently hot to potentially drive significant evaporation from a surviving gas giants and so possibly accrete non-rocky, volatile material \citep{schreiberetal19-1}. White dwarfs at this transition point in their evolution may display evidence of accretion of both rocky and gaseous planets offering an unprecedented opportunity to study the evolution of diverse planetary systems. 

However the quality of the data currently at our disposal severely limits our ability to analyse this system. Without $H$ and/or $K_s$ near-IR photometry the characterization of the thermal flux from the debris disc relies almost exclusively on $WISE$ photometry which could be contaminated by background sources. 
Furthermore, with an average resolution of $\simeq 3$ \AA, the initial ISIS spectroscopy of this star does not reveal any metal absorption lines which, at $\Teff=29\,282$\,K, are expected to be very weak in the optical range. 
Nonetheless, the spectrum reveals emission features of O, Fe and Ca and clearly shows that the \ion{Ca}{II} emission lines are extremely asymmetrical with the red portion of the double-peaked profile being almost absent. This line morphology is closely reminiscent of that of HE\,1349--2305 which shows rapid variability in the line profile indicating a precession period of $1.4\,\pm0.2$\,yr \citep{dennihyetal18-1}. WD\,J2133+2428 is therefore a particularly promising target for spectral variability monitoring as the emission line profile could likely show marked changes on relatively short timescales. 

\subsection{\texorpdfstring{WD\,J221202.88--135239.96}{WD J221202.88-135239.96}}

Similarly to WD\,J0846+5703, WD\,2212--1352 was identified as a gaseous debris disc host via detection of the blended \ion{Mg}{I}/\ion{Fe}{II} emission at $\simeq 5175$\,\AA\, in our initial MagE ID spectrum (Figure\,\ref{Fast_MagE})  

WD\,J2212--1352 is one of the only two He-dominated atmosphere white dwarfs in our sample, the other one being WD\,J0234--0406 described before. However, aside from the main atmospheric composition, the two systems have very little in common and stand as a testament to the diversity found among gas emission line systems.
While WD\,J0234--0406 is one of the coolest white dwarfs in our sample ($\Teff= 13\,454 \pm 200$\,K), WD\,J2212--1352 is among the hottest emission line white dwarfs known, surpassed only by the recently discovered SDSS\,J0006+2858 and WD\,J210034.65+212256.89 \citep{dennihyetal20-1, melisetal20-1}, in addition to WD\,J2133+2428 discussed above.

The lower opacity of a He-dominated atmosphere compared to a H one \citep{camisassaetal17-1} makes detection of metal pollutants easier, and WD\,J2212--1352 and WD\,J0234--0406 appear as the most polluted systems in our sample. However, the nature of this pollution is significantly different in the two objects. 
No H is detected in the atmosphere of WD\,J2212--1352, most likely as a result of the short accretion history of this system. With a cooling age $t_{\mathrm{cool}}\simeq21$\,\,Myr, WD\,J2212--1352 is much younger than WD\,J0234--0406 ($t_{\mathrm{cool}}\simeq281$\,My) and simply would not have had the time to accrete any significant amount of hydrogen \citep{mustilletal18-1, hoskinetal20-1}. 

\begin{figure}
	\includegraphics[width=\columnwidth]{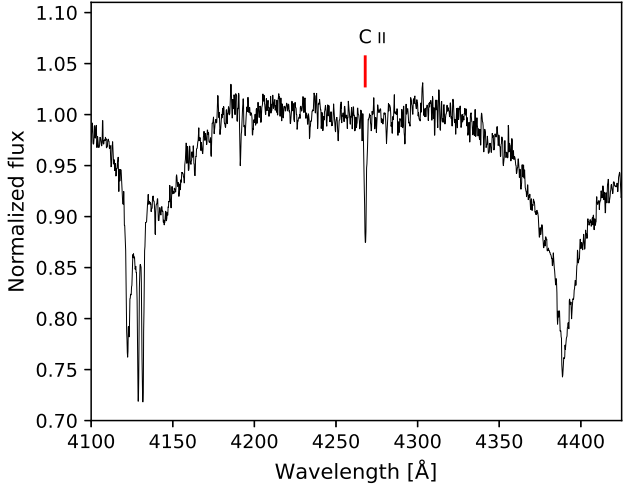}
    \caption{\ion{C}{II} absorption line in the 
    2016-09-28 X-shooter spectrum of  WD\,J2212--1352.}
    \label{C_line}
\end{figure}

Additionally, WD\,J2212--1352 displays clear C absorption (Fig.\,\ref{C_line}), which is not seen in WD\,J0234--0406 (only very few white dwarfs accrete planetary debris that is sufficiently rich in C to be detected at optical wavelengths, e.g. \citealt{juraetal15-1}). Though reconstruction of the bulk chemical composition will require more accurate modelling, we can conclude that the debris about WD\,J2212--1352 likely originated from a body significantly different in composition from that of WD\,J0234--0406.

Though the IR excess about both stars is very similar, corresponding to blackbodies of roughly equal $T_{\textrm{BB}}$ with close $\tau$ values (Table\,\ref{mega_table}), the emission line profiles are significantly different with WD\,J2212--1352 displaying much stronger emission and narrower double peaked profiles. Moreover, the spectrum of WD\,J2212--1352 shows additional emission lines of Fe and O, not detected in WD\,J0234--0406. 
This marked difference is most likely the result of both the different \Teff\ and debris composition of these two stars, making WD\,J2212--1352 and WD\,J0234--0406 perfect examples of the diversity which is starting to emerge among gas disc hosts.

\begin{figure}
    \centering
	\includegraphics[width=\columnwidth]{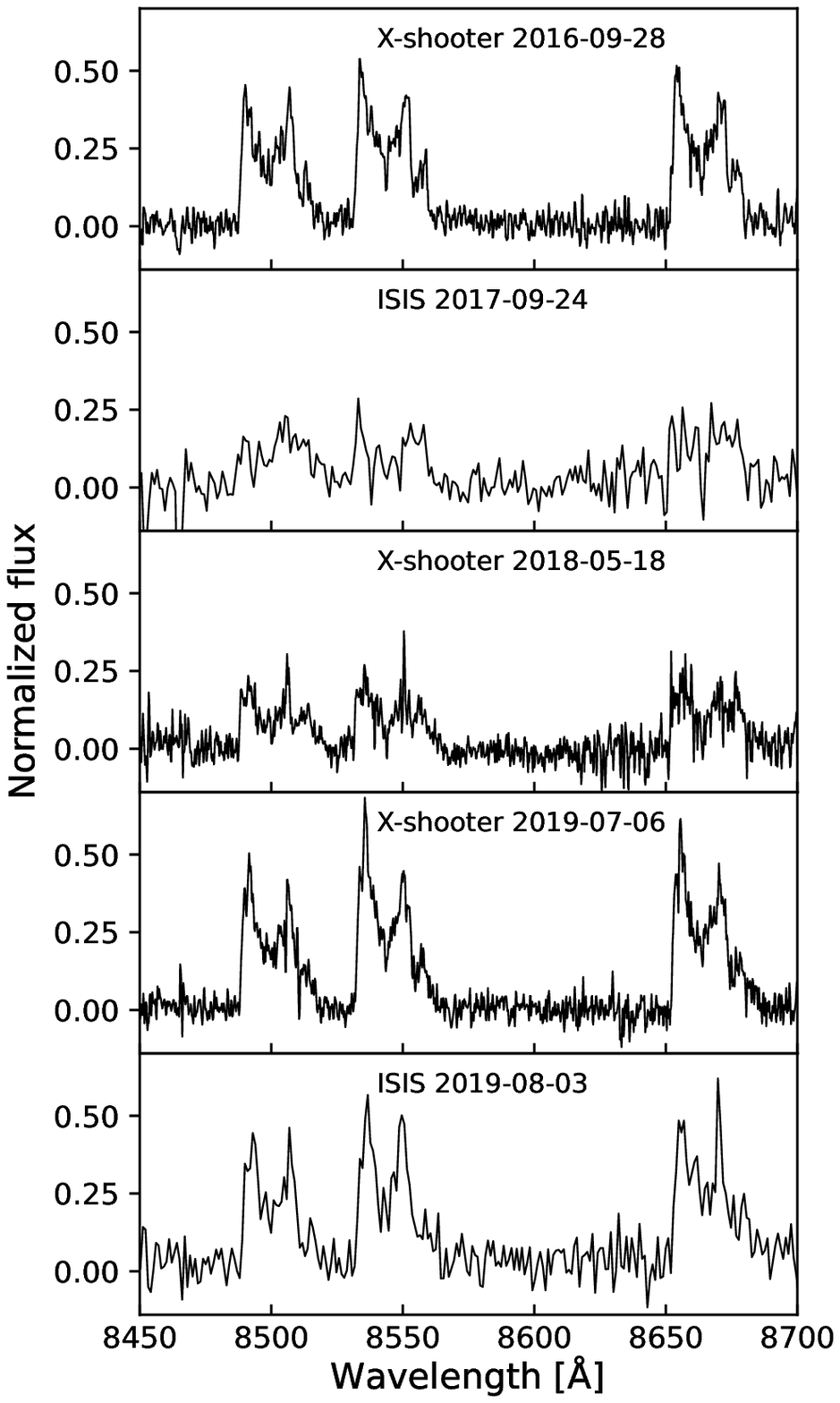}
	\includegraphics[width=0.95\columnwidth]{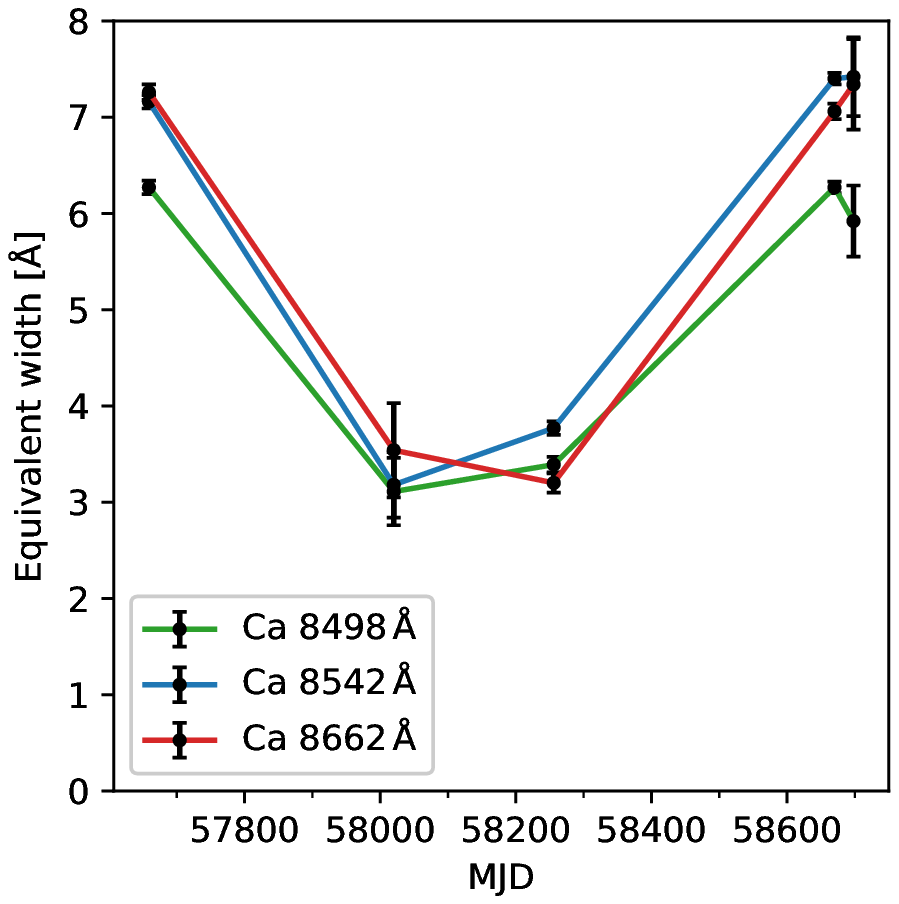}
    \caption{\textit{Top panel}: Different epochs of spectroscopic observations of the \ion{Ca}{II} emission lines in  WD\,J2212--1352.
    \textit{Bottom panel}: Change in the equivalent width of the \ion{Ca}{II} triplet emission lines of WD\,J2212--1352, showing a marked dimming between 2017 and 2018.}
    \label{WDJ2212_epoch}
\end{figure}

To date we have acquired five different epochs of spectroscopic observations of WD\,J2212--1352 covering the \ion{Ca}{II} emission triplet. As shown in Fig.\,\ref{WDJ2212_epoch} the \ion{Ca}{II} lines appear to have undergone significant changes during the course of three years. Between 2016 and 2017 the strength of the \ion{Ca}{II} emission triplet decreased by $\simeq50$\,per\,cent. Similar behaviour has been documented for the white dwarfs WD\,J1617+1620 \citep{wilsonetal14-1}, SDSSJ0845+2257 \citep{wilsonetal15-1}, SDSS\,J1043+0855 \citep{manseretal16-2}, WD\,J034736.69+162409.73  and WD\,J210034.65+212256.89 \citep{dennihyetal20-2}, though with some noteworthy differences.

In the case of WD\,J1617+1620 the emission lines were initially undetectable, only appearing during a brightening event followed by a decline to the ground state over the course of a decade. The authors interpreted the appearance of the gaseous disc signature as a stochastic episode, potentially caused by a collision within the dust disc \citep{wilsonetal14-1}.

WD\,J210034.65+212256.89, underwent a similar brightening event followed by a return to a state with undetectable emission lines, but over a course of only sixty days. This lead the authors to speculate that the appearance of the lines may be a recurring event though no observations showing a return to a bright state have been published so far \citep{dennihyetal20-2, melisetal20-1}.
The emission lines of WD\,J034736.69+162409.73 have also been observed to increase in brightness by $\simeq$\,30\,per\,cent over 300\,days, but to the latest observation of this system there is no evidence of a subsequent dimming \citep{dennihyetal20-2}. 

In contrast, the Ca emission triplet in SDSSJ0845+2257 was initially observed in a bright state in 2004 and by 2008 the EW of the lines had dropped by $\simeq40$\,per\,cent. In subsequent observations taken until 2014 the system remained in this ``low'' state with no significant change in the EW \citep{wilsonetal15-1}. The gaseous emission from the disc around SDSS\,J1043+0855 also appears to vary from a bright state, dropping in brightness by $\simeq$\,40--50\,per\,cent between 2003 and 2010, but then re-brightening by $\simeq$\,30\,per\,cent from 2010 to 2012 \citep{manseretal16-2}.

Similarly to  SDSS\,J1043+0855, the \ion{Ca}{II} emission lines in WD\,J2212--1352 were observed to re-brighten after dropping between 2016 and 2017, returning to their initial strength in 2019. The variability in WD\,J2212--1352 represents the strongest \textit{re}-brightening event of a gaseous debris disc observed to date, but at this stage the data is insufficient to speculate on the cause of this transitory change in flux or on whether this could happen again. Nonetheless it is evident that gaseous debris discs are intrinsically dynamical environments and these recurring changes in emission strength could potentially be the signpost of a stochastic disruption events and/or collisions within the debris disc.

In addition to the changes observed in the strength of the emission lines, comparing the 2016 spectrum of WD\,J2212--1352 with the one obtained in 2019 (Fig.\,\ref{WDJ2212_epoch}), some evidence of morphological changes also emerge. The redshifted portions of the double\-peak profiles appear stronger in 2019 than in 2016 and the lines seem to be transitioning to a more symmetrical morphology. Such evolution could be a hint of precession of a fixed intensity pattern as observed in the archetypal system SDSS\,J1228+1040 \citep{manseretal16-1}, but the data at hand are insufficient to speculate further.

\section{Discussion}
The nine new gaseous debris discs recently announced by \citet{dennihyetal20-2} and \citet[including WD\,J0846+5703]{melisetal20-1}, together with the additional five presented here, tripled the number of these enigmatic evolved planetary systems, to a total of 21. This substantially larger sample of gaseous debris discs allows us to start assessing the properties of these systems as a class rather than on a case-by-case basis. We summarise the key characteristics of the full sample of known gaseous debris disc hosts in Table\,\ref{full_sample}.

The \ion{Ca}{II} triplet emission profiles presented in Figure\,\ref{CaII} show a wide range of morphologies and strengths, with the emission from WD\,J0529--3401, WD\,J2133+2428, and WD\,J2212--1352 showing strong asymmetries. Every \ion{Ca}{II} triplet emitting system that has both multi-epoch observations and asymmetric emission profiles presents morphological variability \citep{gaensickeetal08-1, wilsonetal15-1, manseretal16-1, manseretal16-2, dennihyetal18-1, dennihyetal20-1, melisetal20-1}. Furthermore, two focused studies on SDSS\,J1228+1040 and HE1349--2305 have shown that the morphological evolution of the profiles are periodic in nature, and well described by the precession of a fixed intensity profile in the disc \citep{manseretal16-1, dennihyetal18-1}. The period of this precession can span the range of years to decades from system to system, and this range is supported by theoretical studies that show the effects of general relativistic precession and pressure forces on the gaseous disc \citep{miranda+rafikov18-1}.

While multiple potential mechanisms have been proposed to explain the production of these gaseous components to debris discs, the number of well-monitored and analysed systems to test these scenarios is low. WD\,J0529--3401, WD\,J2133+2428, and WD\,J2212--1352 represent ideal systems (amongst just the systems presented here) for us to follow-up given their asymmetric emission profiles, and the hints of morphological variability, revealed by the multi-epoch observations we have obtained so far.

\afterpage{
\clearpage
\begin{landscape}
\begin{table}
\centering
\caption{\label{full_sample} Known sample of white dwarfs hosting gaseous debris discs in emission ordered by $T_{\textrm{eff}}$. System parameters are given with errors where known, and elements identified in emission are reported. Maximum EW of the \ion{Ca}{II} triplet recorded for each system are reported here. This is an expanded table adapted from \protect\cite{manseretal16-2}.
$^{1}$ \protect\cite{gentilefusilloetal19-1},
$^{2}$ \protect\cite{melisetal20-1}, 
$^{3}$ \protect\cite{farihietal12-1},
$^{4}$ \protect\cite{xuetal14-1},
$^{5}$ This paper,
$^{6}$ \protect\cite{wilsonetal14-1},
$^{7}$ \protect\cite{dufouretal12-1},
$^{8}$ \protect\cite{gaensickeetal07-1},
$^{9}$ \protect\cite{melisetal10-1},
$^{10}$ \protect\cite{manseretal16-2},
$^{11}$ \protect\cite{melis+dufour17-1}, 
$^{12}$ \protect\cite{dennihyetal20-2}, 
$^{13}$ \protect\cite{koesteretal05-2},
$^{14}$ \protect\cite{vossetal07-1},
$^{15}$ \protect\cite{melisetal12-1},
$^{16}$ \protect\cite{wilsonetal15-1},
$^{17}$ \protect\cite{gaensickeetal06-3},
$^{18}$ \protect\cite{gaensickeetal12-1},
$^{19}$ \protect\cite{koesteretal14-1},
$^{20}$ \protect\cite{manseretal16-1},
$^{21}$ \protect\cite{manseretal19-1},
}
\begin{tabular}{lllllllllr}
\hline
WDJ name & Alternate Name & Type & $\log\,g$ & $T_{\textrm{eff}}$ (K) & $M_{\mathrm{WD}}$ (\Msun) & $\tau_{\textrm{cool}}$ (Myr) & Ca\,{\textsc{ii}} EW [\AA] & Detected elements & Reference \\
\hline
WD\,J0147+2339  & WD\,0145+234        & DAZ  & 8.1 (0.1)     & 12720 (1000) & 0.67          & 367 (50)  & 6.7 (0.3)             & Ca & 1, 2 \\
WD\,J0959--0200 & WD\,0956--017       & DAZ  & 8.06 ( 0.03)  & 13280 (20)   & 0.64 (0.02)   & 324 (17)  & 3.0 (1.0)             & Ca & 3, 4 \\
WD\,J1930--5028 &                     & DAZ  & 7.90 (0.05)   & 13306 (500)  & 0.55 (0.03)   & 236 (48)  & 12.48 (0.15)          & Ca, Mg, 5175$^\dagger$ &  1, 5 \\
WD\,J0234--0406 &                     & DABZ & 8.01 (0.03)   & 13454 (200)  & 0.59 (0.02)   & 282 (25)  & 2.39 (0.09)$^\ddagger$ & Ca, 5175$^\dagger$ & 1, 5 \\
WD\,J1617+1620  & WD\,1615+164        & DAZ  & 8.11 (0.08)   & 13520 (200)  & 0.68 (0.05)   & 350 (50)  & 26.38 (0.66)          & Ca & 6 \\
WD\,J0738+1835  & WD\,0735+187        & DBZ  & 8.4 (0.2)     & 13950 (100)  & 0.841 (0.131) & 477 (160) & 19.31 (2.23)          & Ca & 7 \\
WD\,J0846+5703  & SBSS\,0842+572      & DAZ  & 8.02 (0.05)   & 17803 (500)  & 0.63 (0.03)   & 115 (24)  & 33.67 (0.08)          & Ca, Mg, Si, Fe & 1, 2, 5 \\
WD\,J1043+0855  & WD\,1041+091        & DAZ  & 8.124 (0.033) & 17879 (195)  & 0.693 (0.020) & 153 (10)  & 18.89 (0.86)          & Ca, 5175$^\dagger$ & 8, 9, 10, 11 \\
WD\,J0611--6931 & Gaia\,J0611--6931   & DAZ  & 8.2 (0.1)     & 17900 (1000) & 0.74          & 159 (59)  & 47.1 (2.6)            & Ca, O, Si, Na, Mg, 5175$^\dagger$ & 1, 2, 12 \\
WD\,J1352--2320 & HE\,1349--2305      & DBAZ & 8.133 (0.06)  & 18173 (0.06) & 0.673         & 149.4     & 5.2 (0.3)             & Ca & 13, 14, 15\\
WD\,J0644--0352 & Gaia\,J0644--0352   & DBAZ & 8.2 (0.1)     & 18350 (1000) & 0.72          & 151 (59)  & 3.6 (0.75)            & Ca & 1, 2, 12 \\
WD\,J1622+5840  & WD\,1622+587        & DBAZ & 7.8 (0.1)     & 18850 (1000) & 0.62          & 60 (24)   & 6.2 (1.7)             & Ca, O, Fe & 1, 2, 12 \\
WD\,J0845+2257  & Ton\,345            & DBZ  & 8.18 (0.20)   & 19780 (250)  & 0.73 (0.11)   & 122 (44)  & 22.89 (0.69)          & Ca, 5175$^\dagger$ & 16 \\
WD\,J1228+1040  & WD\,1226+110        & DAZ  & 8.150 (0.089) & 20713 (281)  & 0.705 (0.051) & 100 (5)   & 74.22 (0.04)          & Ca, O, Mg, Fe & 9, 17, 18, 19, 20, 21 \\
WD\,J0510+2315  & LAMOST\,J0510+2315  & DAZ  & 8.2 (0.1)     & 21700 (1000) & 0.75          & 80 (35)   & 2.2 (0.9)             & Ca, O, Mg, Fe & 1, 2 \\
WD\,J0347+1624  & SDSS\,J0347+1624    & DAZ  & 8.1 (0.1)     & 21815 (1000) & 0.69          & 60 (29)   & 21.8 (1.23)           & Ca, O, Fe & 1, 2, 12 \\
WD\,J0529--3401 &                     & DAZ  & 8.01 (0.04)   & 23197 (300)  & 0.64 (0.02)   & 34 (6)    & 4.64 (0.04)           & H, Ca, Mg, O, Fe & 1, 5 \\
WD\,J2212--1352 & ATLAS\,J2212--1352  & DBZ  & 7.94 (0.03)   & 24969 (400)  & 0.58 (0.02)   & 21 (2)    & 21.54 (0.08)          & Ca, Mg, Fe & 1, 5 \\
WD\,J0006+2858  & SDSS\,J0006+2858    & DAZ  & 8.0 (0.1)     & 26000 (1000) & 0.64          & 18 (8)    & 47.7 (3.3)            & Ca, O, Fe & 1, 2 \\
WD\,J2100+2122  & Gaia\,J2100+2122    & DAZ  & 8.1 (0.1)     & 26550 (1000) & 0.49          & 29 (12)   & 4.6 (0.4)             & Ca, O, Fe & 1, 2, 12 \\
WD\,J2133+2428  &                     & DAZ  & 7.85 (0.03)   & 29282 (400)  & 0.57 (0.02)   & 10 (1)    & 26.5 (0.6)            & Ca, O, Fe, 5175 & 1, 5 \\

\hline
\multicolumn{9}{l}{$^\dagger$ Mg and Fe blend at $\simeq$\,5175\,\AA\ where either element cannot be definitively identified from this feature alone.}\\
\multicolumn{9}{l}{$^\ddagger$ Strongly contaminated by photospheric absorption feature(s). EW quoted is a lower limit.}\\
\end{tabular}
\end{table}
\end{landscape}
\clearpage
}
\begin{figure}
	\includegraphics[width=\columnwidth]{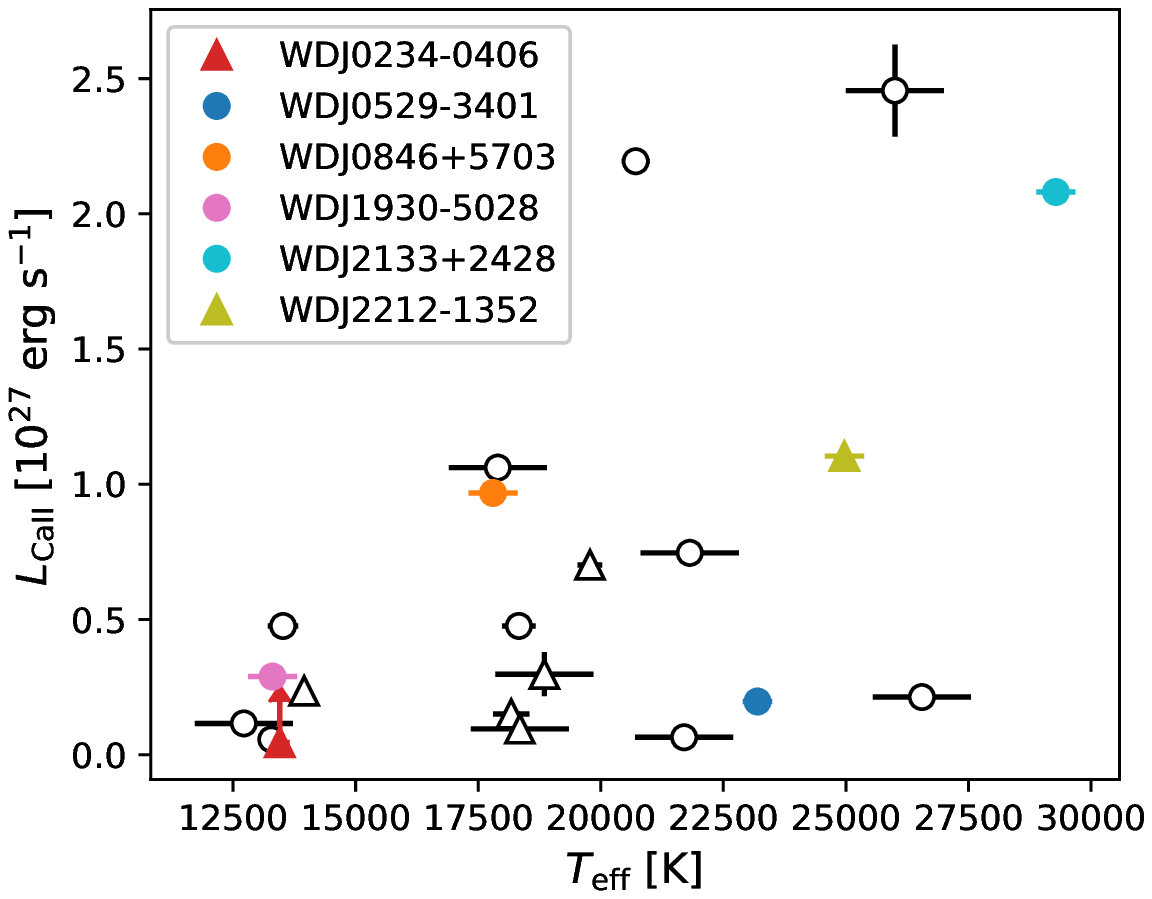}
    \caption{$L_{\ion{Ca}{II}}$ at maximum recorded EW 
    as a function of  \Teff\ for all known gaseous debris disc systems which display Ca emission. H-atmosphere white dwarfs are represented by circles and He-atmosphere white dwarfs with triangles. Previously published systems are shown in black \citep{farihietal12-1,melisetal12-1,dufouretal12-1, wilsonetal14-1,wilsonetal15-1, manseretal16-1, manseretal16-2, dennihyetal20-2,melisetal20-1}, and the six stars presented in this article are shown with different colours. The EW value for WD\,J0234-0406 is a lower limit.}
    \label{all_disc_eqw}
\end{figure}

An interesting relationship to explore is that between the gaseous disc brightness and the \Teff\ of the white dwarf, as it may help to distinguish between different mechanisms responsible for the line emission. For example, emission due to photo--ionisation should be strongly dependent on the temperature of the white dwarf, whereas it should not be so if the temperature in the disc is mostly driven by viscous heating \citep{hartmannetal11-1, hartmannetal16-1}.
However, EW measurements of the \ion{Ca}{II} triplet only represent line strength calculated above the continuum-flattened spectrum, so they do not preserve information on the absolute flux. Therefore, when attempting to compare disc brightness across the broad \Teff\ range spanned by this class of white dwarfs one cannot rely directly on EW values. 
The only information publicly available for all gaseous debris disc hosts published elsewhere is \Teff, $\log g$  and the EW of \ion{Ca}{II} triplet, 
so we opted to define a ``Disc Luminosity Proxy'' ($L_{\ion{Ca}{II}}$) as an indicator of \ion{Ca}{II} triplet luminosity according to the equation:

\begin{equation}
L_{\ion{Ca}{II}}=\mathrm{EW}\,[\angstrom] \times F_{\mathrm{WD}}\times d^{2}/250\,[\angstrom] 
\end{equation}

\noindent where $F_{\mathrm{WD}}$ is the white dwarf flux calculated by integrating a white dwarf model of the appropriate \Teff\ and $\log$\,g between 8450 and 8700\,\AA, and $d$ is the the distance to the white dwarf.

As illustrated in Fig.\,\ref{all_disc_eqw}, the six stars presented in this work uniformly populate the \Teff\  range spanned by the known gaseous debris disc systems, with the discovery of WD\,J2133+2428 extending the upper limit of this distribution by $\simeq3000$\,K. The expanded temperature range populated as well by WD\,J0529--3401, WD\,J2212--1352, SDSS\,J0006+2858 WD\,J210034.65+212256.89 ($\Teff\simeq 26\,000$\,K from \citealt{dennihyetal20-2, melisetal20-1}) allows us to study the effects of \Teff\  on the strengths of the emission lines, as well as their morphology and the species that are detected. WD\,J0529--3401 and WD\,J210034.65+212256.89 also stand out by having, by far, the largest number of detected emission lines. These two systems will be critical for developing quantitative models of the photo-ionised gas disc, using the diagnostic power provided by the multiple transitions of a range of elements. 
So far, such an approach has only been attempted for the purely gaseous disc detected around the white dwarf WD\,J091405.30+191412.25 \citep{gaensickeetal19-1}.

\begin{figure*}
	\includegraphics[width=1.8\columnwidth]{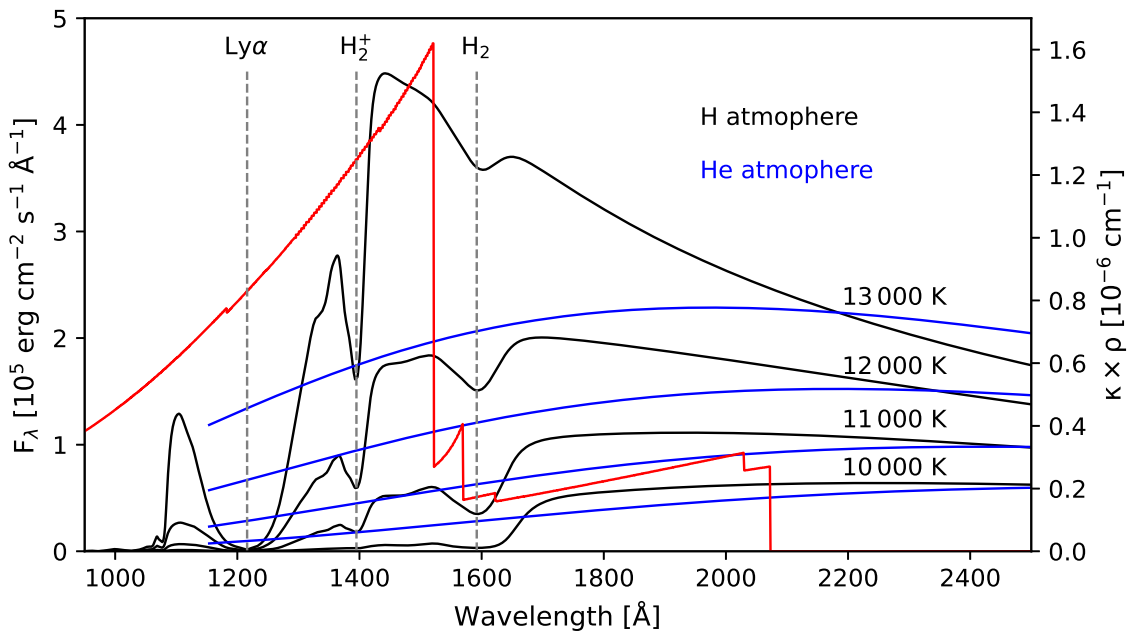}
    \caption{UV fluxes intercepted at a distance of $4.25\times10^{10}$\,cm from the white dwarf for four \Teff\ values ($\log g=8$). H- and He-atmosphere models are displayed with black and blue lines, respectively (the stars discussed in this paper will exhibit in addition some absorption lines of photospheric metals from the accreted debris, which does however not change the illustrative nature of this figure). The quasi-molecular absorption features of H$_2^+$ \citep{koesteretal85-1} and H$_2$ contribute strongly to the suppression of flux at wavelengths $\la1600$\,\AA, in H-atmosphere white dwarfs. Red line: The opacity ($\kappa\rho$ with $\kappa$ the mass absorption coefficient and $\rho$ the density), for a gas of $\rho=5.4\times10^{-12}\mathrm{g\,cm^{-3}}$ and for the abundances of the debris at GD\,362 \citep{zuckermanetal07-1}.}
    \label{opacity}
\end{figure*}

The distribution of known gaseous debris discs as a function of \Teff\ (Fig.\,\ref{all_disc_eqw}) also displays an apparent cut-off at $\simeq12\,500$\,K, though several white dwarfs with dusty debris discs are cooler than that \citep{rocchettoetal15-1}. The gas in the debris discs is heated by the radiation from the white dwarf, and subsequently cools predominantly via emission lines at optical wavelengths \citep{melisetal10-1}.  Modelling the conditions in both gaseous discs composed of rocky-dominated  \citep{kinnear11-1} and volatile-dominated \citep{gaensickeetal19-1} planetary material suggests typical densities in the line-forming regions of $\rho\simeq10^{-12}-10^{-11}\mathrm{g\,cm^{-3}}$. The opacity of this gas is dominated by bound-free cross-sections of metals in the UV (Fig.\,\ref{opacity}). As the white dwarf cools, its UV flux drops. In H-atmosphere white dwarfs, this is exacerbated by the strong absorption of Ly$\alpha$, and the quasi-molecular $H_2^+$ and $H_2$ features at 1400\,\AA\ and 1600\,\AA, respectively \citep{wegner82-1, koesteretal85-1}. He-atmosphere white dwarfs have, compared to their H-analogues, lower UV fluxes because of the absence of line blanketing of the Lyman lines (Fig.\,\ref{opacity}). However even trace amounts of H (likely present in a large fraction of He atmosphere white dwarfs \citealt{koesteretal15-1}) would cause significant UV flux absorption, and re-distribution at other wavelengths. For temperatures $\le12\,000$\,K, the white dwarf flux below $\simeq1600$\,\AA\ quickly drops, resulting in a rapid drop in the heating of the circumstellar gas, and hence a reduction in the emission line fluxes~--~which is consistent with the cut-off of gaseous debris discs around 12\,500\,K.

All white dwarfs with gaseous debris discs also contain trace-metals in their photospheres, as they accrete from the circumstellar material. These metals will imprint narrow absorption lines into the UV spectra of these white dwarfs (see e.g. fig.\,1 of \citealt{gaensickeetal12-1} and fig.\,1 of \citealt{wilsonetal15-1}), which may result in some amount of line blanketing. Whereas this will somewhat (by at most a few per cent) modify the temperature at which the UV flux of a given white dwarf becomes inefficient at photoionising the circumstellar gas, it does not change the conclusion drawn here.
In addition to the observed cut-off at low \Teff, there appears to be a trend of increasing gas disc brightness with increasing \Teff.
This corroborates the idea that the emission line flux correlates with the white dwarf UV luminosity and, as the number of photo-ionising photons increases with higher \Teff, so does the strength of emission lines.
However, there are also a number of outlying systems (WD\,J0529--3401 among them) which display significantly lower emission line flux compared to other white dwarfs of similar \Teff.
This suggests that the \Teff\ of the white dwarf in not the only factor which determines the strength of the emission line and other properties of the disc must also play role. Given the discs are optically thick in the emission lines \citep{gaensickeetal06-3, melisetal10-1}, the gas surface density will have little impact on the total strength of the emission profiles. We therefore hypothesise that both the location of the inner and outer edges of the gaseous material, and  the surface area covered by the gas, are important factors determining the strength of the \ion{Ca}{II} triplet. 
Alternatively, given the variability in emission line strength displayed by many of these white dwarfs, we cannot exclude that these ``low emission'' systems may  have, so far, only been observed during a ``low phase''. 

The \ion{Ca}{II} triplet is often the strongest set of emission lines in the spectra of these stars and it has been used as the defining feature of this class of objects. However, all six systems presented in this article also display \ion{Fe}{II} and/or \ion{Mg}{I} emission at $\simeq$\,5175\,\AA\ (Table\,\ref{all}). Furthermore, WD\,J0846+5703 and WD\,J2212--1352  were both discovered to host a gaseous debris disc from the detection of these emission lines, and only dedicated follow-up observations revealed the \ion{Ca}{II} triplet. Among the other fifteen published systems, nine have an emission feature detected in this region \citep{gaensickeetal06-3, wilsonetal15-1, manseretal16-2, dennihyetal20-2, melisetal20-1}. With over two-thirds of gaseous debris discs showing evidence of an \ion{Fe}{II} or \ion{Mg}{I} $\simeq$\,5175\,\AA\ emission feature, we consider this region to be a useful secondary identifier for the presence of a gaseous debris disc whenever the \ion{Ca}{II} triplet region is not covered spectroscopically.
\section{conclusion}
We have reported observations of of six  white dwarfs which display emission lines from a circumstellar gaseous debris disc, of which five are new discoveries: 
WD\,J0234--0406, WD\,J0529--3401, WD\,J0846+5703, WD\,J1930--5028, WD\,J2133+2428, and WD\,J2212--1352. All objects were identified as candidate debris disc hosts from a marked IR excess in their SEDs, and follow-up spectroscopy revealed Doppler-broadened emission lines, a signature of the presence of a gaseous disc of  planetary debris.
These six systems display very different levels of metal pollution both in terms of number of polluting elements and their nature. Their IR excesses, $\tau$, are comparable to those observed among other white dwarfs with dusty discs, with the exception of WD\,J0846+5703. The IR excess in this latter system is too large to be modelled by the standard optically-thick, geometrically-thin passive disc model, and confirms previous suggestions that the distribution of the circumstellar dust may be more complex than assumed so far.

Finally, we find that the emission lines of these six gaseous discs vastly differ in shape (strength, width and symmetry of the profile), in total number, range of elements (with WD\,J0529--3401 displaying over 50 emission lines from Fe, O, H, Mg and Ca), and in variability behaviour (with WD\,J0846+5703 showing limited change over a time scale of 3\,yr, whereas WD\,J2212--1352 displays 50\,per\,cent  variations in EW over the same time interval).

Together with the recent discoveries of \citet{dennihyetal20-2} and \citet{melisetal20-1}, the sample of gaseous debris discs now includes 21 systems.
The broad range of parameter space spanned by these stars points towards a vast underlying diversity in these  systems. Most notably the white dwarf \Teff\ range from  $\simeq12\,700$\,K  to over 29\,000\,K (WD\,J2133+2428 the hottest white dwarf with a debris disc known to date).

Systems that display emission lines from multiple elements, like WD\,J0529--3401, also represent ideal targets for modelling the gaseous emission profiles akin to the study performed by \cite{gaensickeetal19-1} for WD\,J091405.30+191412.25. This will allow one to place  better constraints the location the the gas, and also measure metal abundances in the disc itself and compare them to those measured from the white dwarf photosphere \citep{gaensickeetal19-1}.

These rich, dynamically active environments still hold a number of secrets, many of which are potentially only unlockable through mid-IR spectroscopy, a type of observation which has been unattainable for years. Debris discs around white dwarf are known to exhibit strong solid-state emission features in their mid-IR spectra \citep{reachetal09-1, juraetal09-1}, which can be directly compared with the elements detected in the photosphere \citep{xuetal14-2}. The advent of \textit{JWST} will enable us to characterize these features and identify the specific chemical compounds which made-up the parent bodies, thus enabling detailed physical modelling of the chemical history of the accreted material.

\section*{Acknowledgements}
We thank the anonymous referee for their constructive review.
We thank P.~G.~Jonker and K.~M.~Lopez for providing the LIRIS observations. We thank Richard Booth for insightful discussion that improved this manuscript. O.T was supported
by a Leverhulme Trust Research Project Grant. 
MRS thanks for support from Fondecyt (grant 1181404) and ANID, -- Millennium Science Initiative Program -- NCN19\_171. PR-G acknowledges support from the State Research Agency (AEI) of the Spanish Ministry of Science, Innovation and Universities (MCIU) and the European Regional Development Fund (FEDER) under grant AYA2017--83383--P.
Based on observations made with the Gran Telescopio Canarias (GTC), installed in the Spanish Observatorio del Roque de los Muchachos of the Instituto de Astrof\'isica de Canarias (IAC), in the island of La Palma. Based on observations made with the William Herschel Telescope operated on the island of La Palma by the Isaac Newton Group of Telescopes in the Spanish Observatorio del Roque de los Muchachos of the Instituto de Astrof\'isica de Canarias. Data for this paper have been obtained under the InternationalTime Programme of the CCI (International Scientific Committee of the Observatorios de Canarias of the IAC). This work is based on observations collected at the European Southern Observatory under ESO programmes 0102.C-0351 and 103.D-0763. Data for this paper have been obtained under the International Time Programme of the CCI (International Scientific Committee of the Observatorios de Canarias of the IAC). Magellan telescope time was granted by NSF’s NOIRLab, through the Telescope System Instrumentation Program (TSIP). TSIP was funded by NSF. Some of the data presented herein were obtained at the W.M. Keck Observatory from telescope time allocated to the National Aeronautics and Space Administration through the agency’s scientific partnership with the California Institute of Technology and the University of California. This work was supported by a NASA Keck PI Data Award, administered by the NASA Exoplanet Science Institute. The Observatory was made possible by the generous financial support of the W.M. Keck Foundation. The authors wish to recognize and acknowledge the very significant cultural role and reverence that the summit of Mauna Kea has always had within the indigenous Hawaiian community. We are most fortunate to have the opportunity to conduct observations from this mountain. BTG, CJM \& TRM were supported by the STFC grant ST/T000406/1. BTG acknowledges support by a Leverhulme Research Fellowship. DV gratefully acknowledges the support of the STFC via an Ernest Rutherford Fellowship (grant ST/P003850/1).  ED acknowledges support by the international Gemini Observatory, a program of NSF’s NOIRLab, which is managed by the Association of Universities for Research in Astronomy (AURA) under a cooperative agreement with the National Science Foundation, on behalf of the Gemini partnership of Argentina, Brazil, Canada, Chile, the Republic of Korea, and the United States of America.

\section{Data availability}
All data underlying this article is publicly available from the relevant observatory archive (see program ID in Table\,\ref{OB_log}) or  will be shared on reasonable request to the corresponding author.

\bibliographystyle{mnras}
\bibliography{aamnem99,aabib1,aabib2,aabib_new,aabib_tremblay}

\appendix 

\section{Additional observations}

    \begin{figure*}
	\includegraphics[width=1.8\columnwidth]{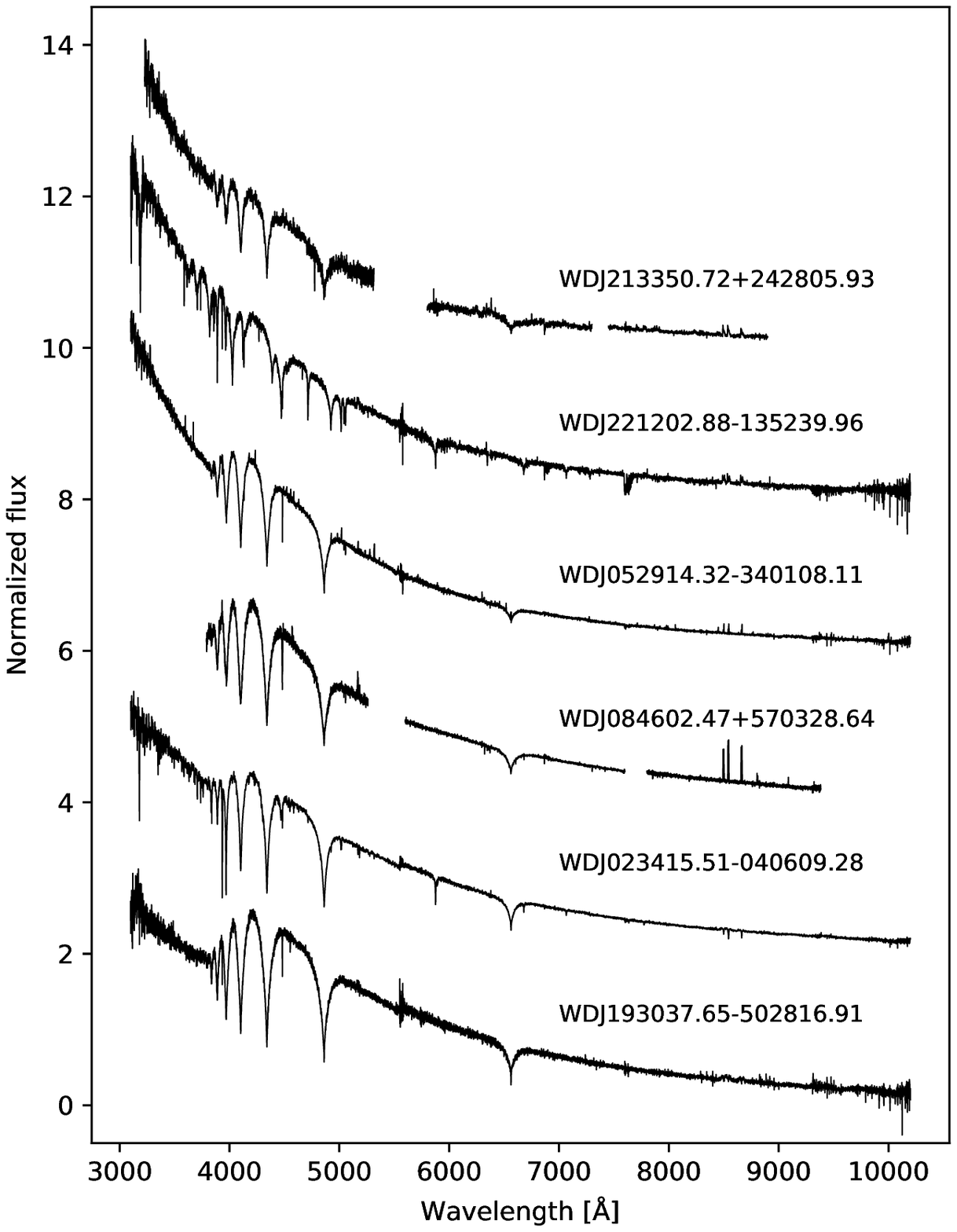}
    \caption{X-shooter and ISIS spectra of our six targets. Spectra have been offset for clarity and arranged in order of decreasing \Teff.}
    \label{all_spec}
\end{figure*}
\begin{figure*}
\centering
	\includegraphics[width=1.7\columnwidth]{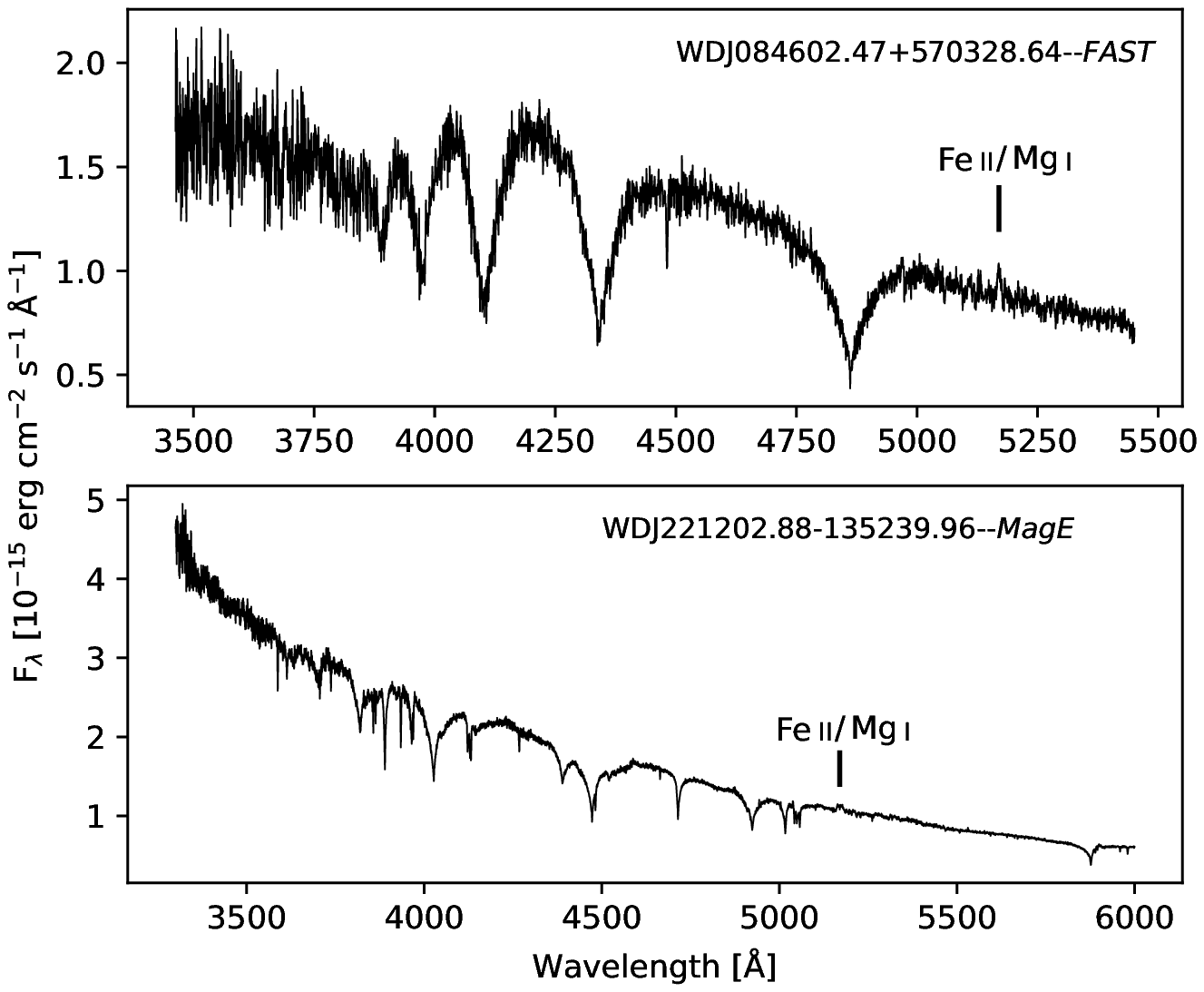}
    \caption{First ID spectroscopy obtained for WD\,J0846+5703 and WD\,J2212--1352.}
    \label{Fast_MagE}
\end{figure*}

\begin{center}
\begin{figure*}
	\includegraphics[width=1.6\columnwidth]{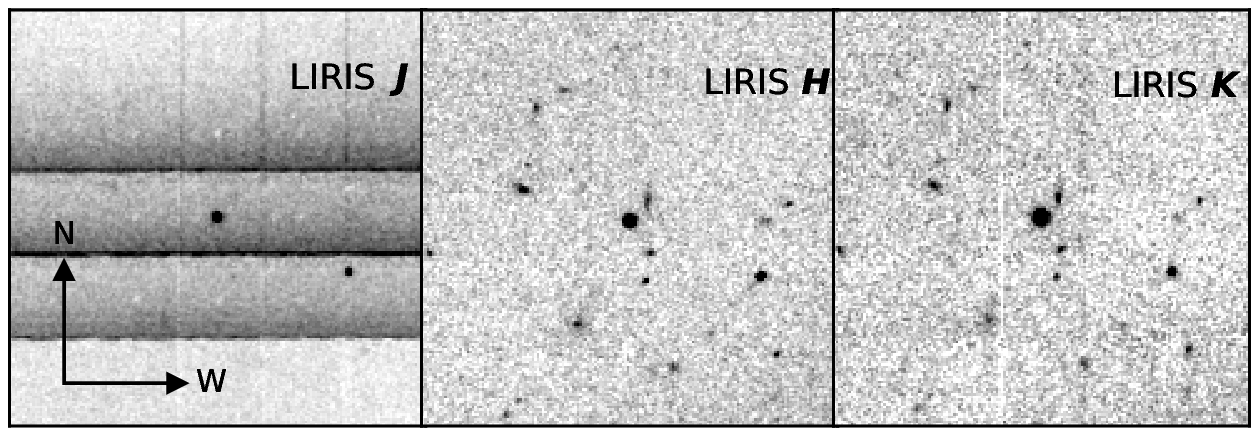}
    \caption{LIRIS images of WD\,0846+5703}
    \label{LIRIS}
\end{figure*}
\end{center}

\section{Emission lines list}
\begin{table*}
\centering
\caption{\label{all} The EWs of all lines identify by eye in the spectra of the six systems presented here. ``:" denotes uncertain or missing element identification.}
\begin{tabular}{llllllll}
\hline
Element    & wavelength [\AA]& \multicolumn{6}{c}{EW [\AA]}                                                \\
           &                 & WD\,J0234 & WD\,J0529 & WD\,J0846 & WD\,J1930   & WD\,J2133 & WD\,J2212\\
\hline
\ion{Fe}{ii}:        & 3277             &                       & 0.21$\pm$0.02     &                  &                     &                  &                  \\
\ion{Fe}\,II    & 4175             &                       & 0.10$\pm$0.02     &                  &                     &                  &                  \\
\ion{Fe}{II}    & 4178             &                       & 0.17$\pm$0.02     &                  &                     & 0.98$\pm$0.23    &                  \\
\ion{Fe}{II}    & 4234             &                       & 0.17$\pm$0.02     &                  &                     & 0.64$\pm$0.16    &                  \\
:        & 4297             &                       & 0.07$\pm$0.02     &                  &                     &                  &                  \\   
:        & 4303             &                       & 0.08$\pm$0.02     &                  &                     &                  &                  \\
\ion{Mg}{I}        & 4352             &                       & 0.19$\pm$0.02     &                  &                     &                  &                  \\
:        & 4377            &                       & 0.07$\pm$0.02     &                  &                     &                  &                  \\
\ion{Fe}{II}    & 4388             &                       & 0.09$\pm$0.02     &                  &                     &                  &                  \\
\ion{Fe}{II} :  & 4417             &                       & 0.07$\pm$0.02     &                  &                     &                  &                  \\
:        & 4492             &                       & 0.14$\pm$0.02     &                  &                     & 1.07$\pm$0.16    &                  \\
\ion{Fe}{II}    & 4520             &                       & 0.28$\pm$0.02     &                  &                     &                  &                  \\
\ion{Fe}{II}        & 4549             &                       & 0.13$\pm$0.02     &                  &                     &                  &                  \\
:    & 4556       &                       & 0.14$\pm$0.02     &                  &                     & 0.79$\pm$0.18    &                  \\
\ion{Mg}{I}        & 4571       &                       & 0.05$\pm$0.02     & 0.37$\pm$0.03    &                     &                  &                  \\
\ion{Fe}{II}    & 4584             &                       & 0.25$\pm$0.02     &                  &                     &                  &                  \\
\ion{Fe}{II}    & 4630             &                       & 0.24$\pm$0.02     &                  &                     &                  &                  \\
\ion{Fe}{II}    & 4924             &                       & 0.31$\pm$0.02     &                  &                     &                  &                  \\
\ion{Fe}{II}    & 5021             &                       & 0.35$\pm$0.02     &                  &                     &                  &                  \\
\ion{Fe}{II}/Mg\,I blend$^{1}$    & $\simeq$\,5175$^{2}$       & 0.28$\pm$0.04$^{3}$   & 0.49$\pm$0.02     & 1.81$\pm$0.04    & 0.70$\pm$0.07       & 2.05$\pm$0.32    & 1.05$\pm$0.04    \\
\ion{Mg}{I}$^{4}$:        & 5184            &                       & 0.07$\pm$0.02     & 0.59$\pm$0.03    &                     &                  &                  \\
\ion{Fe}{II}:           & 5198             &                       & 0.22$\pm$0.02     &                  &                     &                  &                  \\
:        & 5234             &                       & 0.19$\pm$0.02     &                  &                     &                  &                  \\
\ion{Fe}{II}    & 5276             &                       & 0.30$\pm$0.02     &                  &                     &                  & 0.44$\pm$0.03    \\
\ion{Fe}{II}    & 5285             &                       & 0.08$\pm$0.02     &                  &                     &                  &                  \\
\ion{Fe}{II}    & 5316             &                       & 0.77$\pm$0.02     &                  &                     &                  & 0.46$\pm$0.03    \\
\ion{Fe}{II}    & 5363             &                       & 0.16$\pm$0.02     &                  &                     &                  &                  \\
\ion{Fe}{II}    & 5426             &                       & 0.11$\pm$0.02     &                  &                     &                  &                  \\
\ion{Fe}{II}    & 5534             &                       & 0.21$\pm$0.09     &                  &                     &                  &                  \\
\ion{Fe}{II} :  & 5991             &                       & 0.16$\pm$0.03     &                  &                     &                  &                  \\
\ion{Fe}{II}    & 6240             &                       & 0.10$\pm$0.03     &                  &                     &                  &                  \\
:        & 6249             &                       & 0.27$\pm$0.03     &                  &                     &                  &                  \\
\ion{Fe}{II} :   & 6318             &                       & 0.37$\pm$0.03     &                  &                     &                  &                  \\
:        & 6385            &                       & 0.19$\pm$0.04     &                  &                     &                  &                  \\
:        & 6457             &                       & 0.40$\pm$0.03     &                  &                     &                  &                  \\
:        & 6493             &                       & 0.16$\pm$0.03     &                  &                     &                  &                  \\
\ion{Fe}{II}    & 6516             &                       & 0.56$\pm$0.02     &                  &                     &                  &                  \\
\ion{H}{I}      & 6561             &                       & 2.13$\pm$0.03     &                  &                     &                  &                  \\
:        & 7514             &                       & 0.40$\pm$0.03     &                  &                     &                  &                  \\
\ion{Fe}{II}    & 7713             &                       & 0.69$\pm$0.04     &                  &                     & 1.29$\pm$0.21    &                  \\
\ion{O}{I}      & $\simeq$\,7774$^{2}$       &                       & 0.43$\pm$0.03     &                  &                     & 3.81$\pm$0.29    &                  \\
:        & 7865             &                       & 0.46$\pm$0.03     &                  &                     &                  &                  \\
:        & 7918             &                       & 0.20$\pm$0.03     &                  &                     &                  &                  \\
\ion{O}{I}      & 8446             &                       & 1.02$\pm$0.04     &                  &                     & 2.94$\pm$0.51    &                  \\
\ion{Ca}{II}    & 8498             & 1.45$\pm$0.10$^{2,3}$ & 1.58$\pm$0.03     & 9.53$\pm$0.08    & 8.40$\pm$0.16$^{2}$ & 9.26$\pm$0.45    & 6.71$\pm$0.08    \\
\ion{Ca}{II}    & 8542             & 1.45$\pm$0.10$^{2,3}$ & 1.69$\pm$0.04     & 12.50$\pm$0.08   & 8.40$\pm$0.16$^{2}$ & 8.42$\pm$0.93    & 7.63$\pm$0.07    \\
\ion{Ca}{II}    & 8662             & 0.94$\pm$0.09$^{3}$   & 1.37$\pm$0.05     & 11.64$\pm$0.08   & 4.08$\pm$0.14       & 8.82$\pm$0.51    & 7.20$\pm$0.08    \\
\ion{Mg}{I} :    & 8806             &                       & 0.35$\pm$0.03     & 2.95$\pm$0.07    & 1.59$\pm$0.17       &                  & 1.12$\pm$0.10    \\
\ion{Ca}{II}    & 8929             &                       & 0.58$\pm$0.04     &                  &                     &                  &                  \\
:        & 9572             &                       & 1.24$\pm$0.07     &                  &                     &                  &                  \\
\ion{Fe}{ii}:        & 9999             &                       & 4.53$\pm$0.15     &                  &                     &                  &                  \\
\hline
\multicolumn{8}{l}{$^{1}$ \ion{Fe}{II} 5169\,\AA\ and/or \ion{Mg}{I} 5167/5173/5184\,\AA\ contribute to this feature.}\\ 
\multicolumn{8}{l}{$^{4}$ Feature is clearly seperated from the \ion{Fe}{II}/\ion{Mg}{I} blend for WD\,J0529-3401 and WD\,J0846+5703.}\\ 
\multicolumn{8}{l}{$^{2}$ Blended emission features measured as single feature.}\\ 
\multicolumn{8}{l}{$^{3}$ Strongly contaminated by photospheric absorption feature(s). EW quoted is a lower limit.}\\ 
\end{tabular}
\end{table*}

\label{lastpage}
\end{document}